\documentclass[12pt,english]{article}
\usepackage[T1]{fontenc}
\usepackage[latin9]{inputenc}
\usepackage{geometry}
\usepackage{array}
\usepackage{rotating}
\usepackage{float}
\usepackage{multirow}
\usepackage{amsmath, mathrsfs}
\usepackage{amssymb}
\usepackage{graphicx}
\usepackage{setspace}
\usepackage{esint}
\usepackage[authoryear]{natbib}
\usepackage{color}
\usepackage{titlesec}
\usepackage{authblk}

\titleformat*{\section}{\large\bfseries}
\titleformat*{\subsection}{\normalfont\bfseries}
\titleformat*{\subsubsection}{\normalfont\bfseries}
\titleformat*{\paragraph}{\normalfont\bfseries}
\titleformat*{\subparagraph}{\normalfont\bfseries}

\makeatletter

\providecommand{\tabularnewline}{\\}
\renewcommand{\baselinestretch}{1.8} 
\newcommand*{\GOWL}{GOWL}

\makeatother

\usepackage{babel}
\begin{document}

\title{Estimating Individualized Treatment Rules for Ordinal Treatments}

\author[1]{\vspace{-2 ex} Jingxiang Chen}
\author[4]{Haoda Fu}
\author[4]{Xuanyao He}
\author[1,2]{Michael R. Kosorok}
\author[1,2,3,*]{Yufeng Liu\vspace{-2 ex}}

\affil[1]{Department of Biostatistics \vspace{-2 ex}}
\affil[2]{Department of Statistics and Operations Research \vspace{-2 ex}}
\affil[3]{Department of Genetics \vspace{-2 ex}}
\affil[ ]{University of North Carolina at Chapel Hill  \vspace{-2 ex}}
\affil[4]{Eli Lilly and Company \vspace{-2 ex}}
\affil[*]{Email: yfliu@email.unc.edu \vspace{-5 ex}}

\date{}

\maketitle{
\renewcommand{\baselinestretch}{1.3}

\noindent
{\bf SUMMARY}: Precision medicine is an emerging scientific topic for disease treatment and prevention that takes into account individual patient characteristics. It is an important direction for clinical research, and many statistical methods have been recently proposed. One of the primary goals of precision medicine is to obtain an optimal individual treatment rule (ITR), which can help make decisions on treatment selection according to each patient's specific characteristics. Recently, outcome weighted learning (OWL) has been proposed to estimate such an optimal ITR in a binary treatment setting by maximizing the expected clinical outcome. However, for ordinal treatment settings, such as individualized dose finding, it is unclear how to use OWL. In this paper, we propose a new technique for estimating ITR with ordinal treatments. In particular, we propose a data duplication technique with a piecewise convex loss function. We establish Fisher consistency for the resulting estimated ITR under certain conditions, and obtain the convergence and risk bound properties. Simulated examples and two applications to datasets from an irritable bowel problem and a type 2 diabetes mellitus observational study demonstrate the highly competitive performance of the proposed method compared to existing alternatives.


}

\newpage
\pagenumbering{gobble} 

\pagenumbering{arabic}

\section{Introduction}

In clinical research, precision medicine is a medical paradigm that promotes personalized health care to individual patients. Its recent development originates from the fact that treatment effects can vary widely from subject to subject due to individual level heterogeneity.
For example, \citet{ellsworth_breast_2010} found that women whose CYP2D6 gene has a certain mutation are not able to metabolize Tamoxifen efficiently, and this makes them an improper target group for this therapy.
In this way, one of the primary goals for precision medicine is to establish rules so that patients level characteristics can be used directly to find optimal treatments \citep{mancinelli_pharmacogenomics:_2000,tania_simoncelli_paving_2014}. Recent literature indicates that statistical machine learning tools can be useful in building such rules. However, the primary focus has been on the binary treatment case, and the ordinal setting has not been explored. Ordinal treatments are commonly seen in practice. For example, some drugs used to treat the same disease can be ranked by their medicinal strengths. Multiple doses of the same treatment can be viewed as ordinal. However, the dose-response relationship is usually discussed from a population perspective in practice \citep{friedman_fundamentals_2010}. In precision medicine, it is desirable to pursue the dose level that is best suited for each individual patient. In this paper, we develop a statistical learning model which can properly handle optimal treatment detection for both binary and ordinal treatment scenarios.

Various novel quantitative methods have been proposed in the statistical learning literature to estimate ITRs. For example, one group of methods aims to construct easily interpretable results by using tree-based methods to explore heterogeneous treatment effects \citep{su_subgroup_2009,su_interaction_2011,lipkovich_strategies_2014, laber_tree-based_2015}.
Another group of methods focuses on establishing a scoring system to evaluate patients' benefits from certain treatments \citep{zhao_effectively_2013}. 
However, these two groups of methods do not propose any optimization function from which the optimal treatment solution can be found. As an alternative, \citet{qian_performance_2011} proposed a value function of the average reward that patients receive from their assigned treatments so that the rule discovery process is transformed into an optimization problem. \citet{zhang_robust_2012} developed inverse probability of treatment weights to robustly estimate such value functions, \citet{fan_concordance-assisted_2016} developed a robust rank regression method to estimate a concordance function for individual treatment regime detection, and \citet{zhao_estimating_2012} proposed outcome weighted learning (OWL) to transform the rule detection problem into a weighted classification problem. In particular, the OWL approach uses a hinge loss function to replace the original 0-1 loss function in \citet{qian_performance_2011}, and thus the corresponding computation becomes feasible. Furthermore, \citet{chen_dose_2016} adjusted the OWL method to find the best dose when treating dose as a continuous variable. 

Although \citet{zhao_estimating_2012} offers clear ideas on how the ITR can be estimated, it still has some challenges in practice.
The first challenge is that OWL's ITR estimate might be suboptimal when some patient rewards are less than zero. In this setting, a global minimization of the loss function cannot be guaranteed since the objective function is no longer convex. If one chooses to manually shift all of the rewards to be positive, the estimated ITR tends to retain what is actually assigned (\citet{zhou_residual_2015}). This phenomenon can become more severe when the sample size is small and the covariate dimension is large. 
To alleviate this problem, \citet{zhou_residual_2015} recently proposed residual weighted learning. However, their resulting object function is non-convex, and consequently, global minimization is still not guaranteed.


An ordinal treatment, a categorical treatment with a defined order to its categories, can be different from nominal treatment and continuous dose in precision medicine. On one hand, an ordinal treatment can give more restrictions on treatment effect estimate when compared with nominal treatments; on the other hand, it is not appropriate to simply consider an ordinal treatment as a continuous variable because the labels do not contain information about difference scales between each two treatment levels. In that case, the discussion remains valuable that how to extend the objective function of OWL to solve the ITR estimation problem for ordinal treatments. Such an extension is non trivial in practice. This is because the objective function of standard OWL maximizes the average reward by adjusting only the observations where the optimal treatment is identical to the actually assigned treatment. In other words, it ignores how different the actual assigned treatment is from the optimal treatment, which leads to information loss. Several methods have been proposed to consider such differences among treatments in the standard ordinal classification learning framework. One previously developed idea in statistical learning is the data duplication strategy introduced by \citet{ling_ordinal_2006} and \citet{cardoso_learning_2007}. This strategy borrows the idea from proportional odds cumulative logistic regression (\citet{agresti_wiley_2014}), which restricts the estimated boundaries not to cross with each other. Furthermore, the ordinal response is relabeled as a binary variable and duplicated in the covariate data to generate a higher dimensional sample space. Then, an all-at-once model is fitted in the transformed sample space to produce a corresponding ranking rule for the original response.
Although such data duplication methods are shown to be effective in solving complex ordinal  classification problems, it remains unclear how this idea can be utilized in OWL to help find the optimal ITR among multiple ordinal treatments.

Motivated by the discussion above, in this paper, we propose a new method called generalized outcome weighted learning (\GOWL).
Specifically, our first contribution is to create a new objective function for  ITR estimation based on the value function definition in \citet{qian_performance_2011} through making use of the data duplication idea. We then formulate the optimal ordinal treatment rule detection problem into an aggregation of several optimal binary treatment rule detection subproblems. Furthermore, considering that each subproblem corresponds to a level of the ordinal treatment, we prevent estimated decision boundaries from the subproblems intersecting with each other to circumvent contradictory results. The second contribution of the paper is to modify the loss function in \citet{zhao_estimating_2012} to maintain convexity regardless of whether the value of the reward is positive or negative.
This loss function enables \GOWL\ to penalize the treatments corresponding to negative reward values properly to avoid the rewards shift problem previously described.

To estimate the optimal individual treatment rule in the new optimization problem, we provide an efficient algorithm using the primal-dual formulation. Moreover, we show that our method achieves Fisher consistency under mild conditions, which means that the true optimal treatment will be reached if the entire population is used. In addition, we prove that the estimated intercepts of the decision functions are monotonic along the treatment level, which will make the decision boundaries interpretable in practice. We also show that  the proposed method with the Gaussian kernel has the asymptotic convergence rate of $n^{-1/2}$ for a well-separated data set under the geometric noise condition (\citet{steinwart_fast_2007}).

The remainder of the paper is organized as follows. In Section 2, we review the OWL method and then explain how the modified loss function for \GOWL\ works under the binary treatment setting. In Section 3, we first illustrate how \GOWL\ works for the ITR estimate in the ordinal treatment setting based on the necessary background information for the data duplication method.
Then, we develop an efficient algorithm to solve the corresponding optimization problem. In Section 4, we establish the statistical learning properties of \GOWL, including Fisher consistency, excess risk bound, and convergence rates. Simulated data examples are used in Section 5, and two applications to an irritable bowel syndrome problem and a type 2 diabetes mellitus observational study are provided in Section 6. We then provide some discussions and conclusions in Section 7.

\section{Generalized Outcome Weighted Learning for Binary Treatments}

In this section, we give a brief review of OWL and its corresponding optimization problem. Motivated by the limitations of OWL, we propose a generalized version of OWL for the binary treatment case using a modified loss function.


\subsection{Outcome Weighted Learning}

Suppose that we collect the data from a two-arm clinical
study where the binary treatment is denoted by $A\in\mathcal{A}=\{-1,1\}$.
We assume that the patients'
prognostic results are represented by an $n$ by $p$ matrix
$X\in\mathcal{X}$, where $\mathcal{X}$ denotes the prognostic space, $n$
is the number of patients enrolled, and $p$ corresponds to the number
of measured prognostic variables. We also use a bounded random variable
$R$ to represent the clinical outcome reward and assume a larger
$R$ value is more desirable. Note that $R$ can depend on both $X$ and $A$. Under this framework, the ITR is a mapping
from $\mathcal{X}$ to $\mathcal{A}$. According to \citet{qian_performance_2011},
the goal of an optimal ITR is to find the mapping $\mathcal{D}=\mathcal{D}^{*}$
such that{\small{}
\begin{equation}
\mathcal{D}^{*}(X)=\underset{\mathcal{D}}{\arg\min}\left\{ E\left(\frac{R\cdot I(A\neq\mathcal{D}(X))}{P(A|X)}|X,\mathcal{D}\right)\right\}, \label{eq:0-1_loss}
\end{equation}
}where $P(A|X)$ is the prior probability of treatment $A$ for $X$. Note that $P(A|X)=P(A)$ under the independence assumption between $A$ and $X$. Furthermore, the expectation operation in (\ref{eq:0-1_loss}) is conditional on $X$ and $\mathcal{D}$. From now on, we will omit the conditional part of the expectation to simplify the expressions. To estimate the optimal treatment rule $\mathcal{D}^*$, one needs to obtain a classifier function $f(x)$ such that $\mathcal{D}(x)=\text{sign}(f(x))$. Thus, we have the following two indicator functions equivalent to each other{\small{}
\begin{equation}
I(A \neq \mathcal{D}(X)) = I(A \cdot f(X)\leq0).
\end{equation}
}To alleviate the non-deterministic polynomial-time (NP) computational
intensity (\citet{feldman_agnostic_2010}) in (\ref{eq:0-1_loss}),
\citet{zhao_estimating_2012} proposed OWL by replacing the 0-1 loss above with the hinge loss used in the Support Vector Machine (SVM, \citet{cortes_support-vector_1995}) together with a regularization term to control model complexity. As a consequence, the regularized optimization problem becomes a search for the decision rule $f$ which
minimizes the objective function{\small{}
\begin{equation}
\frac{1}{n}\sum_{i=1}^{n}\frac{r_{i}}{P(a_{i}|x_i)}\left[1-a_{i}f(x_{i})\right]_{+}+\lambda||f||^{2}, \label{eq:owl_obj}
\end{equation}
}where $(x_{i},a_{i},r_{i}); i=1,\cdots,n$, is a realization of $(X,A,R)$ with $a_i\in \{-1,1\}$,
the function $\left[u\right]_{+}=\max(u,0)$ denotes the positive
part of $u$, $||f||^{2}$ is the squared $L_{2}$ norm of $f$
and $\lambda$ is the tuning parameter used to control the model complexity and avoid overfitting. Notice that to maintain the convexity of the objective function, OWL requires all rewards to be non-negative.

 In practice, when there are negative rewards, one can shift them by a constant to ensure positiveness. \citet{zhou_residual_2015} noted that such a constant shift process for the rewards may lead to suboptimal estimates. In particular, they noted that the optimal treatment estimates tend to be the same as the random treatments that are originally assigned. 
This situation can be further illustrated by a toy example as follows. Suppose we have two intervention groups (treatment and placebo) and two patients both being assigned to the treatment group and receiving rewards of $-10$ and $10$, respectively.
Such results imply that the first patient may not benefit from the treatment due to the corresponding negative feedback. If we follow the reward shift idea as mentioned above and add $15$ to both  rewards, then the model will probably draw an incorrect conclusion that both patients benefit from the treatment since both shifted rewards are positive. Another controversy of this rewards-shift strategy comes from the fact that there are an infinite number of constants one can choose for the shift. Different shift constants can lead to different coefficient estimates when the decision rule $f$ has a certain parametric or nonparametric form in problem (\ref{eq:owl_obj}). To solve this problem, we propose a generalized OWL in Section 2.2 which does not require rewards to be positive.


\subsection{Generalized Outcome Weighted Learning}

\begin{figure}[h!]
\begin{centering}
\includegraphics[scale=0.4]{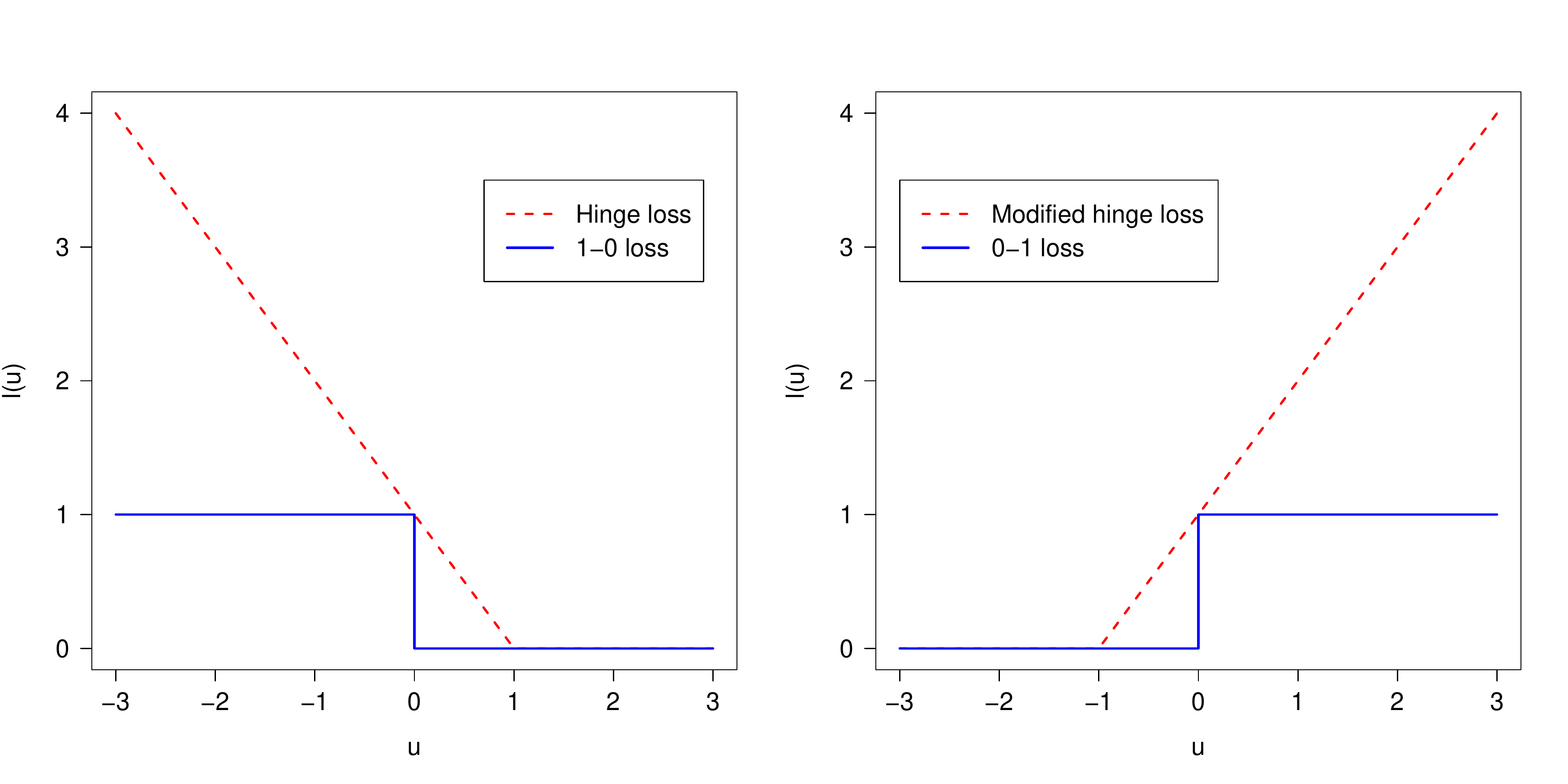}
\par\end{centering}

\protect\caption{Standard hinge loss $l_{1}(u)=\left[1-u\right]_{+}$ versus 1-0 loss (left) and modified
hinge loss $l_{2}(u)=\left[1+u\right]_{+}$ versus 0-1 loss (right). The modified
hinge assigns large loss values to those observations whose estimated treatment
rule matches the actual treatment assigned.\label{fig:Modified-Hinge-Loss}}
\end{figure}

For problem (\ref{eq:owl_obj}), note that the OWL objective function is convex only when all of the rewards are non-negative and such a restriction
could make OWL suboptimal when there are negative rewards, as discussed earlier. To remove such a restriction, we first consider reformulating the minimization problem (\ref{eq:0-1_loss}) into two pieces as {\small{}
\begin{equation}
\underset{\mathcal{D}}{\arg\min} E\left\{  \frac{|R|}{P(A|X)}\left[I(R\geq0)I\left(A \neq \mathcal{D}(X)\right)+I(R<0)I\left(A=\mathcal{D}(X)\right)\right]\right\}. \label{eq:modified-0-1}
\end{equation}
}Note that (\ref{eq:modified-0-1}) is equivalent to (\ref{eq:0-1_loss}) because the term $\frac{R\cdot I(R<0)}{P(A|X)}$ is free of $\mathcal{D}(X)$. Similar to the discussion in Section 2.1, we can rewrite the optimization problem in (\ref{eq:modified-0-1}) as follows, with $\mathcal{D}(X)=\text{sign}(f(X))$:{\small{}
\begin{equation}
\underset{\mathcal{D}}{\arg\min} E\left\{  \frac{|R|}{P(A|X)}\left[I(R\geq0)I\left(A\cdot f(X) \leq 0\right)+I(R<0)I\left(A\cdot f(X)>0\right)\right]\right\}. \label{eq:modified-Af}
\end{equation}
}Furthermore, to alleviate the computational intensity of solving (\ref{eq:modified-Af}), we use a modified loss function to be minimized with the population form expressed as {\small{}
\begin{equation}
E\left\{ \frac{|R|}{P(A|X)}\left[I(R\geq0)\left[1-Af(X)\right]_{+}+I(R<0)\left[1+Af(X)\right]_{+}\right]\right\}. \label{eq:modified OWL_population}
\end{equation}
}Here the ITR $\mathcal{D}$ in (\ref{eq:modified-0-1}) is the sign function of the decision rule $f$ in (\ref{eq:modified OWL_population}) by definition. Therefore, the corresponding empirical sum on the training data becomes {\small{}
\begin{equation}
\sum_{i=1}^n\left\{ \frac{|r_i|}{P(a_i|x_i)}\left[I(r_i\geq0)\left[1-a_if(x_i)\right]_{+}+I(r_i<0)\left[1+a_if(x_i)\right]_{+}\right]\right\}. \label{eq:modified OWL}
\end{equation}
}Note that the loss in (\ref{eq:modified OWL}) has two parts according to the sign of $r_i$. For observations with positive rewards, we use $r_i$
as their weights for the corresponding loss function and penalize the mis-classification by the standard
hinge loss function $l_1(u)=\left[1-u\right]_{+}$ (see the left panel in Figure \ref{fig:Modified-Hinge-Loss} for how the hinge loss approaches the 1-0 loss).
This part is identical to the hinge loss in OWL. However, for observations
with negative rewards, we use $-r_i$ as their weights
instead and employ a modified hinge loss $l_2(u)=\left[1+u\right]_{+}$
(see the right plot in Figure \ref{fig:Modified-Hinge-Loss} for how the modified hinge loss approaches the 0-1 loss) which
assigns a larger loss to the observations whose estimated treatment
$f(x_i)$ matches the observed treatment $a_i$. As a consequence, the modified
loss function in (\ref{eq:modified OWL}) is piecewise convex in terms of $a_if(x_i)$ (\citet{tsevendorj_piecewise-convex_2001}).
Therefore, a global optimization of the objective function could be
guaranteed when standard convex optimization algorithms are applied. One advantage of using the modified hinge loss is that the observed rewards are no longer required to be positive so that the problem caused by the non-unique reward shift can be circumvented. In addition,
one can see that the loss function reduces to the standard hinge
loss when all $r_i>0$. As a remark, we note that \citet{laber_adaptive_2011} previously used a similar surrogate loss for construction of the adaptive confidence intervals for the test error in classification.

\section{Generalized Outcome Weighted Learning for Ordinal Treatments}

In this section, we discuss how to extend \GOWL\ from binary treatments to ordinal treatments. For problems with multiple ordinal treatments, it is important to utilize the ordinal information. To this end, we borrow the idea of data duplication in standard ordinal classification and develop our new procedure for \GOWL\ with ordinal treatments.

\subsection{Classification on Ordinal Response with Data Duplication}

For an ordinal response problem, suppose each observation vector is $\left(x_{i}^T,y_{i}\right)$ where $i=1,\cdots,n$, the predictor $x_{i}$ contains $p$ covariates, and the response $y_{i}\in\{1,\cdots,K\}$. \citet{cardoso_learning_2007} proposed a data duplication technique to address this problem. To apply this idea, one first needs to generate a new data set written
as $({x_{i}^{(k)}}^T,y_{i}^{(k)})$, where $x_{i}^{(k)}=(x_{i}^T, e_k^T)^T$,
$y_{i}^{(k)}=\text{sign}\left(y_{i}-k\right)$, $e_k^T$ is a $K-1$ dimensional row vector whose $k$th element is 1 while others are zeros, and $k=1,\cdots,K-1$. Thus, $y_i^{(k)}$ defines a new binary response indicating $1,\cdots, k$ versus $k+1,\cdots,K$. Here the $\text{sign}(x)$ function is defined to be $1$ when
$x>0$ and $-1$ otherwise. Then, the goal of the classification
method is to find a surrogate binary classifier $f(x^{(k)})$ to minimize $\sum_{i=1}^{n}\sum_{k=1}^{K-1}l(y_{i}^{(k)},f(x^{(k)}))+J(f)$,
where $l(\cdot)$ is the pre-defined loss and $J(f)$ is a penalty
term. Once these $f(x_{i}^{(k)})$
are obtained  for $k=1,\cdots,K-1$, then the predicted rule $\hat{\mathcal{D}}(x_i)$ for the original
ordinal outcome $y_i$ can be calculated by $\hat{\mathcal{D}}(x_i)=\sum_{k=1}^{K-1}I(f(x_{i}^{(k)})>0)+1$, where $I(\cdot)$ is the indicator function. 

\subsection{Generalized Outcome Weighted Learning}

Now consider an extended version of clinical data $(X,A,R)$
in Section 2 with $X$ and $R$ the same as before but with $A$ being an ordinal treatment
with $A\in\mathcal{A}=\{1,\cdots,K\}$. In contrast to standard multicategory treatment scenarios, the $K$ categories of treatments are ordered in a way that $1$ and $K$ are most different, For example, these treatments may represent different discrete dose levels with $A=1$ being the lowest dose and $A=K$ being the highest dose. Similar to Section 3.1, we define the duplicated random set $\left(X^{(k)},A^{(k)},R^{(k)}\right)$ with its $i$th realization defined as ${x_i^{(k)}}=(x_i^T, e_k^T)^T$,
$a_i^{(k)}=\text{sign}\left(a_i-k\right)$, and $r_i^{(k)}=r_i$ for $k=1,\cdots,K-1$. According to the value function definition from \citet{qian_performance_2011}, we let $P^{\mathcal{D}_{k}}$ denote the conditional distribution
of $(X,A,R)$ on $A^{(k)}=\mathcal{D}(X^{(k)})$. Then, with the
duplicated data set and a map $\mathcal{D}$ from each $X^{(k)}$ to
$\{-1,1\}$ for $k=1,\cdots,K-1$, we propose a new conditional expected reward to be maximized as follows:{
\begin{eqnarray}
\sum_{k=1}^{K-1}E\left(R|A^{(k)}=\mathcal{D}(X^{(k)}),X\right) & = & \sum_{k=1}^{K-1}\int R\frac{dP^{\mathcal{D}_{k}}}{dP}dP\nonumber \\
 & = & \sum_{k=1}^{K-1}\int R\frac{I(A^{(k)}=\mathcal{D}(X^{(k)}))}{P(A|X)}dP\nonumber \\
 & = & \sum_{k=1}^{K-1}E\left(\frac{R\cdot I(A^{(k)}=\mathcal{D}(X^{(k)}))}{P(A|X)}\right).\label{eq:value_function_gowl}
\end{eqnarray}
}Similar to \citet{qian_performance_2011} and \citet{zhao_estimating_2012},
we refer to (\ref{eq:value_function_gowl}) as
the value function of $\mathcal{D}$ and denote it by
$\mathcal{V}(\mathcal{D})$. In this way, the optimal map $\mathcal{D}^{*}$
is defined as{
\begin{eqnarray}
\mathcal{D}^*=\underset{\mathcal{D}}{\arg\min}\sum_{k=1}^{K-1}E\left(\frac{R\cdot I(A^{(k)}\neq\mathcal{D}(X^{(k)})}{P(A|X)}\right).
\label{eq:min_gowl}
\end{eqnarray}
}Once the map $\mathcal{D}$ is estimated, one can obtain the
corresponding ITR estimate of $X$ by using $\hat{\mathcal{D}}(X)=\sum_{k=1}^{K-1}I(f(X^{(k)})>0)+1$.

Notice that optimal treatment estimation through (\ref{eq:min_gowl}) can be effective when the treatment is ordinal due to the way it utilizes the ordinality information. In particular, the new minimization problem considers the distance between the estimated optimal treatment and the actually assigned treatment
by counting the number of mismatches between each $\mathcal{D}(X^{(k)})$ and each $A^{(k)}$ for $k=1,\cdots,K-1$.
In the extreme case when a certain subject has an extremely large positive reward value, the estimated $\mathcal{D}(X^{(k)})$
would be likely to match $A^{(k)}$ for all $k=1,\cdots,K-1,$ which results in $\hat{\mathcal{D}}(X)=A$. In contrast, it may imply that the actually assigned treatment is suboptimal when the reward outcome takes a small value. Some of the estimated $\mathcal{D}(X^{(k)})$ will not match the observed $A^{(k)}$ as the estimated rule approximates the global minimizer of (\ref{eq:min_gowl}).

To alleviate the computational intensity of the minimization problem in (\ref{eq:min_gowl}), we replace the 0-1 loss with
the modified loss in (\ref{eq:modified OWL}) proposed in Section 2.2 and add the model complexity penalty term to avoid overfitting. Thus, the new objective function on $(x_{i}^{(k)},a_{i}^{(k)},r_{i}^{(k)})$ becomes
\begin{equation}
\sum_{i=1}^{n}\sum_{k=1}^{K-1}\frac{\left|r_{i}\right|}{P(a_{i}|x_i)}\left[I(r_{i}\geq0)\left[1-a_{i}^{(k)}f(x_{i}^{(k)})\right]_{+}+I(r_{i}<0)\left[1+a_{i}^{(k)}f(x_{i}^{(k)})\right]_{+}\right]+\lambda||f||^{2},\label{eq:GOWL object}
\end{equation}
where $x_{i}^{(k)}$ is the $k$th duplication of the $i$th original
subject and $f(x_{i}^{(k)})$ is the corresponding binary classifier.
Similarly, the predicted optimal ITR of the $i$th subject $x_{i}$
can be obtained by $\hat{\mathcal{D}}(x_{i})=\sum_{k=1}^{K-1}I(f(x_{i}^{(k)})>0)+1$. In Section 4, we show that our method is Fisher consistent in the sense that the estimate matches $\underset{\mathcal{D}}{\arg\max}E(R|X,\mathcal{D})$ asymptotically under certain mild conditions.



\subsection{Computational Algorithm for \GOWL}

We now introduce our algorithm to solve  (\ref{eq:GOWL object}). Due to the convexity of the objective function in (\ref{eq:GOWL object}),
we generalize the primal-dual method \citet{vazirani_approximation_2013}
used in SVM to estimate the classifier $f(x_{i}^{(k)})$.
Starting from (\ref{eq:GOWL object}), by introducing a series of slack
variable $\xi_{i}^{(k)}$ and $\psi_{i}^{(k)}$ for all  observations
$i=1,\cdots,n$ and all duplicates $k=1,\cdots,K-1$, we rewrite the minimization in (\ref{eq:GOWL object}) by minimizing the following objective function with respect to $f$ and all slack variables, \begin{equation}
\sum_{i=1}^{n}\sum_{k=1}^{K-1}\frac{\left|r_{i}^{(k)}\right|}{P(a_{i}|x_i)}\left[I(r_{i}^{(k)}\geq0)\xi_{i}^{(k)}+I(r_{i}^{(k)}<0)\psi_{i}^{(k)}\right]+\lambda||f||^{2},\label{eq:GOWL_reform}
\end{equation}
with $\xi_{i}^{(k)}\geq0,\psi_{i}^{(k)}\geq0,\xi_{i}^{(k)}\geq1-a_{i}^{(k)}f(x_{i}^{(k)})$,
and $\psi_{i}^{(k)}\geq1+a_{i}^{(k)}f(x_{i}^{(k)})$. 


Next, we discuss how to solve (\ref{eq:GOWL_reform}) for  the linear case in Section 3.3.1 and the non-linear case in Section 3.3.2.

\subsubsection{Linear Decision Function Estimation}

Suppose that the decision function $f(x_{i}^{(k)})$ above is a linear
function of $x_{i}^{(k)}$ with the slope $\tilde{\beta}$ and an
intercept $\tilde b$, i.e. $f(x_{i}^{(k)})=\left[x_{i}^{(k)}\right]^{T}\tilde{\beta}+\tilde b$. Before introducing the algorithm, we express $f(x_{i}^{(k)})=\left[x_{i}^{(k)}\right]^{T}\tilde{\beta}+\tilde b=x_{i}\beta+b_{k}$ by denoting $x_{i}^{(k)}=(x_{i}^T,e_k^T)^T$, where $e_k^T$ is a $K-1$ dimensional row vector whose $k$th element is 1 while others are zeros.  Note that $\tilde{\beta}^T=(\beta^T,b_1-\tilde b,\cdots,b_{K-1}-\tilde b)$. In other words, the decision function on the duplicated covariate
set $x_{i}^{(k)}$ can also be understood as a varying intercept function
of $x_{i}$, i.e. $f(x_{i}^{(k)})=g(x_{i})+b_{k}$.
On one hand, such a form of the decision function constructs $K-1$ parallel boundaries
in the original sample space to avoid contradicting classifying
results. On the other hand, for the ordinal treatment scenario, it is usually desirable to have the $K-1$ intercepts monotonic along the treatment group in terms of the interpretation,
i.e. $b_{i}<(>)b_{i+1}$ for all $i=1,\cdots,K-2$ when $K\geq3$. We show in Section 4 that \GOWL\ enjoys such a property under a reasonable condition. When the assumption of parallel linear boundaries becomes too strong, one can use nonlinear learning techniques to achieve more flexible boundaries as in Section 3.3.2.

To solve (\ref{eq:GOWL_reform}) with a linear decision function, we plug the expression of $f(x_{i}^{(k)})$ above back into (\ref{eq:GOWL_reform}) and reparamatrize
the formula as:{
\[
\underset{\tilde \beta, \xi, \psi}{\min} \left\{\frac{1}{2}||\tilde \beta||^{2}+C\sum_{i=1}^{K-1}\sum_{k=1}^{K-1}\frac{\left|r_{i}^{(k)}\right|}{P(a_{i}|x_i)}\left[I(r_{i}^{(k)}\geq0)\xi_{i}^{(k)}+I(r_{i}^{(k)}<0)\psi_{i}^{(k)}\right]\right\},
\]
}with {\small{}$\xi_{i}^{(k)}\geq0,\psi_{i}^{(k)}\geq0,\xi_{i}^{(k)}\geq1-a_{i}^{(k)}f(x_{i}^{(k)})$,
}  {\small{}$\psi_{i}^{(k)}\geq1+a_{i}^{(k)}f(x_{i}^{(k)})$}, and ($\xi$,$\psi$) denote all slack variables. By introducing the Lagrange multipliers, we can derive the Lagrange function
for the primal problem as:{
\begin{eqnarray*}
L_{P} & = & \frac{1}{2}||\tilde \beta||^{2}+C\sum_{i=1}^{n}\sum_{k=1}^{K-1}\frac{\left|r_{i}^{(k)}\right|}{P(a_{i}|x_i)}\left[I(r_{i}^{(k)}\geq0)\xi_{i}^{(k)}+I(r_{i}^{(k)}<0)\psi_{i}^{(k)}\right]\\
 & - & \sum_{i=1}^{n}\sum_{k=1}^{K-1}\mu_{i}^{(k)}\xi_{i}^{(k)}-\sum_{i=1}^{n}\sum_{k=1}^{K-1}\nu_{i}^{(k)}\psi_{i}^{(k)}-\sum_{i=1}^{n}\sum_{k=1}^{K-1}\alpha_{i}^{(k)}\left[a_{i}^{(k)}f(x_{i}^{(k)})+\xi_{i}^{(k)}-1\right]\\
 & - & \sum_{i=1}^{n}\sum_{k=1}^{K-1}\eta_{i}^{(k)}\left[-a_{i}^{(k)}f(x_{i}^{(k)})+\psi_{i}^{(k)}-1\right].
\end{eqnarray*}
}The corresponding dual problem can be derived by taking partial derivatives with respect to $(\tilde \beta,\xi,\psi)$
and simplifying the results using the Karush\textendash Kuhn\textendash Tucker
conditions (\citet{kuhn_nonlinear_1951}). Then, the dual problem becomes maximizing $L_D$ with respect to the slack variables $\{\alpha_{i}^{(k)},\eta_{i}^{(k)}; i=1,\ldots,n; k=1,\ldots, K-1\}$,  where
\begin{eqnarray*}
L_{D} & = & \sum_{i=1}^{n}\sum_{k=1}^{K-1}\alpha_{i}^{(k)}+\sum_{i=1}^{n}\sum_{k=1}^{K-1}\eta_{i}^{(k)}-\frac{1}{2}\sum_{i=1}^{n}\sum_{k=1}^{K-1}\sum_{j=1}^{n}\sum_{h=1}^{K-1}\alpha_{i}^{(k)}\alpha_{j}^{(h)}a_{i}^{(k)}a_{j}^{(k)}\left(\left[x_{i}^{(k)}\right]^{T}\cdot\left[x_{j}^{(h)}\right]\right)\\
 & - & \frac{1}{2}\sum_{i=1}^{n}\sum_{k=1}^{K-1}\sum_{j=1}^{n}\sum_{h=1}^{K-1}\eta_{i}^{(k)}\eta_{j}^{(h)}a_{i}^{(k)}a_{j}^{(k)}\left(\left[x_{i}^{(k)}\right]^{T}\cdot\left[x_{j}^{(h)}\right]\right),
\end{eqnarray*}
with $0\leq\alpha_{i}^{(k)}\leq\frac{C\cdot r_{i}^{(k)}}{P(a_{i}|x_i)}I(r_{i}^{(k)}\geq0),0\leq\eta_{i}^{(k)}\leq\frac{C\cdot r_{i}^{(k)}}{P(a_{i}|x_i)}I(r_{i}^{(k)}<0)$,
and $\sum_{i=1}^{n}(\alpha_{i}^{(k)}-\eta_{i}^{(k)})a_{i}^{(k)}=0$.
Note that the parameters in the dual problem above can be solved by applying standard quadratic
programming with linear constrains (\citet{nocedal_numerical_2006}). Furthermore, the slope estimate can be obtained via $\hat{\tilde \beta}=\sum_{i=1}^{n}\sum_{k=1}^{K-1}(\hat{\alpha}_{i}^{(k)}a_{i}^{(k)}\text{sign}(r_{i}^{(k)}\geq0)x_{i}^{(k)})$.
The intercept vector $\left\{ b_{1},\cdots,b_{K-1}\right\} $ 
can be estimated by plugging $\hat{\tilde \beta}$ back into the original
maximization in (\ref{eq:GOWL object}) and solving a standard
linear programming problem with linear constraints (\citet{vazirani_approximation_2013}).
Because there are $2n(K-1)$ parameters in the dual problem above, with a finite $K$, the computational
complexity of  (\ref{eq:GOWL object}) is the same as that of the standard primal-dual problem in the SVM.

\subsubsection{Nonlinear Decision Function Estimation}

Section 3.3.1 solves (\ref{eq:GOWL_reform}) for the linear case. However, in practice, the linear assumption can be too strong for some problems. To make our model more flexible, we 
perform nonlinear learning by applying the kernel learning approach in Reproducing
Kernel Hilbert Spaces (RKHS). Kernel learning in RKHS is flexible and has
achieved great successes in many nonlinear learning studies \citep{wahba_spline_1990, scholkopf_learning_2001,shawe-taylor_kernel_2004,hastie_elements_2011}.

Under the binary treatment case, we can show by the Representer Theorem (\citet{kimeldorf_correspondence_1970}) that under some regularity conditions, the decision function on the data $(x_{i}^{(1)}, a_{i}^{(1)}, r_{i}^{(1)})$ can be written in the form that $f(x_{i}^{(1)}) = \sum_{j=1}^{n}k(x_i,x_j)c_{j}+\tilde b$, where $k(\cdot,\cdot)$ is the standard kernel function associated with the RKHS $\mathcal{H}$. When the treatment is extended into an ordinal variable, we need to define an extended version of the kernel function on the duplicated covariates $x_{i}^{(k)}$ to construct the decision function. In particular, we have $f(x_{i}^{(k)})=\sum_{j=1}^{n}\sum_{h=1}^{K-1}\tilde k(x_{i}^{(k)},x_{j}^{(h)})\tilde c^{(h)}_{j}+\tilde b$, where $\tilde k(\cdot,\cdot)$ is the extended kernel function with the definition $\tilde k(x_{i}^{(k)},x_{j}^{(h)})=k(x_i,x_j)+e_{k}^T\cdot e_{h}$, and $e_{k}$ is defined as in  Section 3.3.1. Similar discussions were made in \citet{ling_ordinal_2006} and \citet{cardoso_learning_2007}. According to the newly defined extended kernel, $f(x_{i}^{(k)})$ can be rewritten as $\sum_{j=1}^{n}k(x_i,x_j)c_{j}+b_k$, where $c_j=\sum_{h=1}^{K-1}\tilde c_j^{(h)}$ and $b_k=\sum_{j=1}^n \tilde c_j^{(k)}+\tilde b$. One can tell from the new  $f(x_{i}^{(k)})$ expression that due to the conversion of the ordinal problem into a big binary problem, the corresponding decision boundaries in the kernel-induced feature space are guaranteed not to cross with each other. Consequently,  the sets $\{f(x^{(k)})<0\}$ for $1 \leq k \leq K-1$ produce more flexible noncrossing boundaries for the $K$ ordinal treatments in the original space.


Given the expression of $f$ with respect to the kernel representation, we can follow similar Lagrange optimizer steps as before to obtain the generalized primal-dual formula. We can derive  the dual problem of maximizing $L_D$ with respect to all slack variables, where  {
\begin{eqnarray*}
L_{D} & = & \sum_{i=1}^{n}\sum_{k=1}^{K-1}\alpha_{i}^{(k)}+\sum_{i=1}^{n}\sum_{k=1}^{K-1}\eta_{i}^{(k)}-\frac{1}{2}\sum_{i=1}^{n}\sum_{k=1}^{K-1}\sum_{j=1}^{n}\sum_{h=1}^{K-1}\alpha_{i}^{(k)}\alpha_{j}^{(h)}a_{i}^{(k)}a_{j}^{(k)}\tilde  k\left(x_{i}^{(k)}, x_{j}^{(h)}\right)\\
 & - & \frac{1}{2}\sum_{i=1}^{n}\sum_{k=1}^{K-1}\sum_{j=1}^{n}\sum_{h=1}^{K-1}\eta_{i}^{(k)}\eta_{j}^{(h)}a_{i}^{(k)}a_{j}^{(k)}\tilde k\left(x_{i}^{(k)}, x_{j}^{(h)}\right),
\end{eqnarray*}
}with $0\leq\alpha_{i}^{(k)}\leq\frac{C\cdot r_{i}^{(k)}}{P(a_{i}|x_i)}I(r_{i}^{(k)}\geq0),0\leq\eta_{i}^{(k)}\leq\frac{C\cdot r_{i}^{(k)}}{P(a_{i}|x_i)}I(r_{i}^{(k)}<0)$, and $\sum_{i=1}^{n}(\alpha_{i}^{(k)}-\eta_{i}^{(k)})a_{i}^{(k)}=0$.
After the dual coefficients are estimated, the decision function can be written as $
f(x_{i}^{(k)})=\sum_{j=1}^{n}\sum_{h=1}^{K-1}\tilde k(x_{i}^{(k)},x_{j}^{(h)})(\hat{\alpha}_{j}^{(h)}a_{j}^{(h)}\text{sign}(r_{j}^{(h)}\geq0))$. 

To implement the quadratic programming in the dual problems above, we use the open source package CVXOPT based on the Python programming in practice. 

\section{Statistical Learning Theory}

In this section, we show Fisher consistency of the estimated ITR, the monotonic property of the intercepts, consistency and convergence rate of the risk
bound for the estimated ITR using \GOWL. We define some essential notation before getting into the details. First, we define the risk associated with 0-1 loss in (\ref{eq:modified-0-1}) as{\small{}
\begin{equation}
\mathcal{R}(f)=\sum_{k=1}^{K-1}\mathcal{R}^{(k)}\left(f\right)=E\left\{ \sum_{k=1}^{K-1}\frac{R}{P(A|X)}I\left(A^{(k)}\neq\text{sign}\left(f(X^{(k)})\right)\right)\right\}, \label{eq:01risk}
\end{equation}
}where $\mathcal{R}^{(k)}\left(f\right)=E\left[\frac{R}{P(A|X)}I\left(A^{(k)}\neq\text{sign}\left(f(X^{(k)})\right)\right)\right]$ for $k=1,\cdots,K-1$
and $f(X^{(k)})$ is an ITR associated decision function. According to (\ref{eq:01risk}), we define its Bayes
risk as $\mathcal{R}(f^{*})=\inf_{f}\left\{ \mathcal{R}(f)|f:\mathcal{X}\rightarrow\mathbb{R}\right\} $ and the corresponding optimal ITR as $\mathcal{D}^*(X)=\sum_{k=1}^{K-1}I(f^*(X^{(k)})>0)+1$.
Correspondingly, we define the $\phi-$risk associated with the surrogate loss in (\ref{eq:modified OWL_population})  as 
{\small{}
\begin{equation}
\mathcal{R}_{\phi}(f)=\sum_{k=1}^{K-1}\mathcal{R}_{\phi}^{(k)}\left(f\right)=E\left\{ \sum_{k=1}^{K-1}\frac{|R|}{P(A|X)}\left[\phi\left(A^{(k)}f(X^{(k)}),R\right)\right]\right\}, \label{eq:surrogaterisk}
\end{equation}
}where {\small{}$\mathcal{R}_{\phi}^{(k)}\left(f\right)=E\left[\frac{|R|}{P(A|X)}\phi\left(A^{(k)}f(X^{(k)}),R\right)\right]$}
and $\phi\left(u,r\right)=I(r\geq0)[1-u]_{+}+I(r<0)[1+u]_{+}$. We
also define the minimal $\phi-$risk as $\mathcal{R}_{\phi}(f_{\phi}^{*})=\inf_{f}\left\{ \mathcal{R}_{\phi}(f)|f:\mathcal{X}\rightarrow\mathbb{R}\right\}$ and the corresponding surrogate optimal ITR as $\mathcal{D}^*_{\phi}(X)=\sum_{k=1}^{K-1}I(f_{\phi}^*(X^{(k)})>0)+1$.
Furthermore, we assume that the number of treatment levels $K$ is finite in the following discussions. All the details of theorem proofs are included in the Supplementary Materials.

\subsection{Fisher Consistency}

Recall that the optimal ITR always corresponds to the treatment that can produce the best expected clinical reward, i.e. $\mathcal{D}^{*}(x)=\underset{k\in\mathcal{A}}{\arg\max}\left[E(R|X=x,A=k)\right]$.
To derive Fisher consistency, we need to show that by using the suggested loss $\phi$ to replace the 0-1 loss, the surrogate optimal 
ITR $\mathcal{D}_{\phi}^*(x)$ matches $\mathcal{D}^{*}(x)$. We divide the process into two steps: first, we show in Lemma
4.1 that $\mathcal{D}_{\phi}^*(x)=\mathcal{D}^{*}(x)$ in the binary treatment case.
Second, the conclusion can be generalized into the ordinal treatment problems under
an additional assumption in Theorem 4.2.

\paragraph{Lemma 4.1}

When $A\in\{1,2\}$, for any measurable function $f$, we
have $\mathcal{D}_{\phi}^*(x)=I\left(f_{\phi}^*(X^{(1)})>0\right)+1=\mathcal{D}^{*}(X),$
where $f_{\phi}^*$ is the minimizer of $\mathcal{R}_{\phi}(f)$ in (\ref{eq:surrogaterisk})
with $K=2$.

\paragraph*{}

To prove Lemma 4.1, one can show that the minimizer
$f_{\phi}^*$ should be within the range of $[-1,1]$ and then we can show $\text{sign}(f_{\phi}^*)=\text{sign}(E\left[R|A=2\right]-E\left[R|A=1\right])$. 

\paragraph{Theorem 4.2}

When $A\in \{1,\cdots,K\}$ and $K$ is an integer greater than 2, we have $\mathcal{D}_{\phi}^*(x)=\sum_{k=1}^{K-1}I(f_{\phi}^*(X_{i}^{(k)})>0)+1=\mathcal{D}^{*}(X)$ under the assumption that $E(R|X,A>k)>E(R|X,A\leq k)\quad\text{if and only if}\quad \mathcal{D}^{*}(X)\geq k$
for $k=1,\cdots,K-1$, where $f_{\phi}^*$ is a measurable function that minimizes $\mathcal{R}_{\phi}(f)$ in  (\ref{eq:surrogaterisk}).

\paragraph*{}

\begin{figure}[H]
\begin{centering}
\includegraphics[scale=0.35]{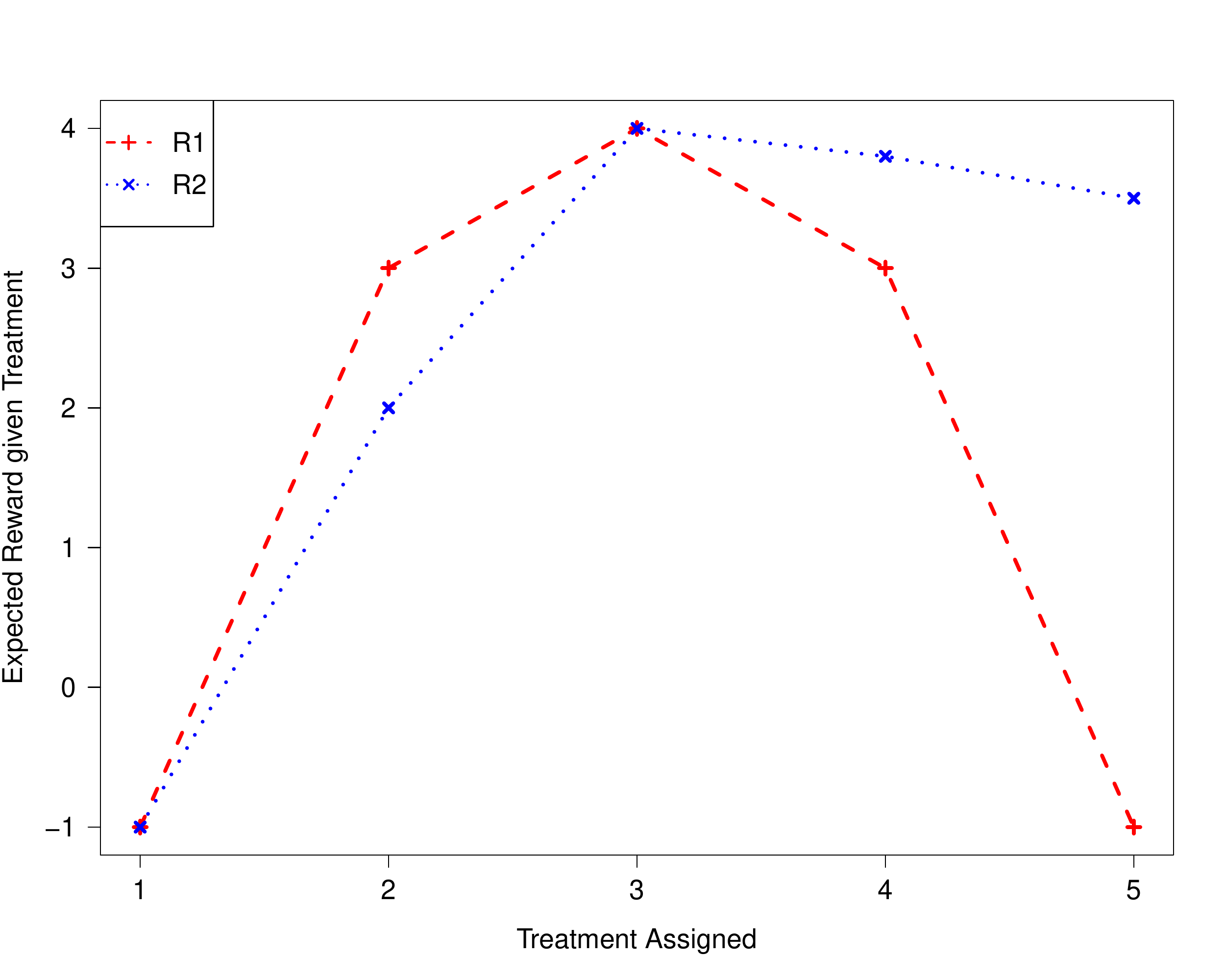}
\par\end{centering}

\protect\caption{Examples when the assumption holds and fails for Theorem 4.2. In this case, $\mathcal{D}^{*}(X)=3$
and the assumption in Theorem 4.2 holds for curve R1 but fails for
curve R2. The assumption of the modified duplication strategy that $R^{(k)}=R\cdot I(A\in\{k,k+1\})$
holds for both curves.\label{fig:The-Assumption-on}}
\end{figure}

To show Theorem 4.2, we start from the conclusion in Lemma 4.1 and obtain $\mathcal{D}^{*}(X)$
by summing all binary decision functions across $k=1,\cdots,K-1$. The assumption
on $E(R|X)$ in Theorem 4.2 is necessary when one needs to accumulate all $f_{\phi}^*(X^{(k)})$ correctly to reach $\mathcal{D}^{*}(X)$. Essentially, this assumption requires
the reward curve decreases at a similar rate when the treatment is away from
the optimal one at both sides of its peak (see the R1 curve in Figure \ref{fig:The-Assumption-on}). According to this assumption, each binary surrogate classifier $I(f_{\phi}^*(X^{(k)})>0)$ matches the corresponding optimal binary classifier $I(f^*(X^{(k)})>0)$ in each binary subproblem. We would like to point out that even when the assumption fails in real applications,
Fisher consistency could still be guaranteed by modifying the data
duplication strategy into $R^{(k)}=R\cdot I(A\in\{k,k+1\})$.
The modified strategy uses partial data in each binary treatment
subproblem so that we only need the reward curve to be monotonically
decreasing when the assigned treatment moves away from the true optimal treatment $\mathcal{D}^{*}(X)$.
Note that the modified duplication strategy uses subsets of data and may work well for large sample problems. In particular, it is well suited for the cases where there is a sufficient sample size within each treatment group.

\subsection{Monotonic Boundary}

In Section 3, we discussed that the decision
function $f(X^{(k)})$ can be expressed as $g(X)+b_{k}$ for both linear and non-linear cases.
The following theorem shows that the intercepts $b_{k}$ for $k=1,\cdots,K-1$ can have the
monotonic property under certain assumptions so that the resulting rule has no contradiction. Note that it is only meaningful to consider the monotonic property of the intercepts when $K\geq3$.

\paragraph{Theorem 4.3}

If we write the decision function as $f(X^{(k)})=g(X)+b_{k}; k=1,\cdots,K-1$,
and assume that the signs of $E\left[R|A=k\right]$ are the same for
$k=1,\cdots,K$, then the optimal solution
$(g,b)$ for minimizing the $\phi$-risk $\mathcal{R}_{\phi}(f)$ has monotonic ${b}$ values. In particular, we have $b_{k}>b_{k+1}$ ($b_{k}<b_{k+1}$) for $k=1,\cdots,K-2$ when $E\left[R|A=k\right]>0$ $ (<0)$ for $k=1,\cdots,K$.

\paragraph*{}


To understand the condition in Theorem 4.3, note that the value of $E[R|A=k]$ is the average benefit patients receive from taking the treatment $k$. Violating the conditions in Theorem 4.3 could destroy the monotonic order of $b$. For example, when $E\left[R|A=m\right]$
for certain $m\in\{2,\cdots,K-1\}$ is observed to be negative while all the other $E[R|A=k]$ are
positive, no patient will be assigned with the treatment $m$ as the optimal treatment and the corresponding $b$ would not be monotonic. 

\begin{figure}[H]
\begin{centering}
\includegraphics[scale=0.47]{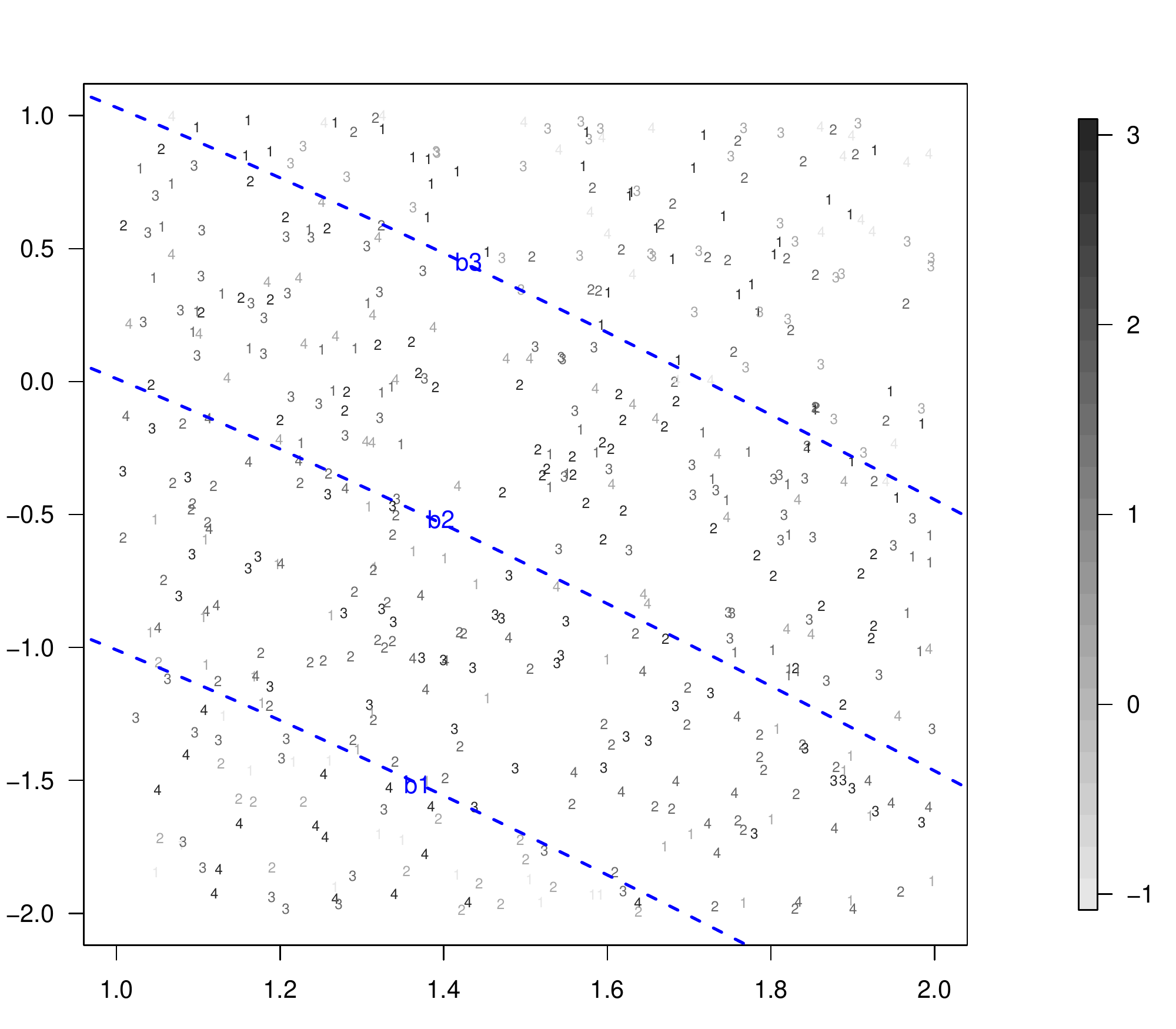}
\par\end{centering}

\protect\caption{A simulation example explaining how the monotonic property works. In this case, there are two covariates and four treatment levels where the numbers represent the actually assigned treatments. The gray-scale of the numbers indicates the clinical outcome value and a darker color means a larger reward (see the gray-scale strip). The dashed lines indicate how the optimal ITR boundaries split the input space. When $E\left[R|A=2\right]$ reduces to a certain negative value that has a large magnitude, the margin between the estimated $b_1$ and $b_2$ boundaries would decrease to zero and then the monotonic property no longer holds. \label{fig:Mono}}
\end{figure}

To further illustrate the condition in Theorem 4.3, Figure \ref{fig:Mono} demonstrates a simulated example with two covariates and four treatment levels where the numbers represent the actually assigned treatments. The gray-scale of the numbers indicates the clinical outcome value and a darker color means a larger reward  (see the gray-scale strip). The dashed lines indicate how the optimal ITR boundaries split the input space into four regions where the optimal treatment rule changes  from $\mathcal{D}^*(x)=1$ in the top right area to $\mathcal{D}^*(x)=4$ in the bottom left. Starting with all positive $E[R|A=k]$, if we decrease $E[R|A=2]$  while keeping the other $E[R|A=k]$ values constant, the margin between $b_1$ and $b_2$ will be narrower. Such a change indicates that a smaller proportion of the population will be assigned $A=2$ as the optimal treatment. In the extreme case where $E[R|A=2]$ is negative and small enough compared with the other two treatments, the boundaries of $b_1$ and $b_2$ will overlap, violating the monotonic property. Under this circumstance, the rewards can contradict the ordinality of the treatments.

Finally, we would like to emphasize that Theorem 4.3 only presents a sufficient condition for the monotonicity of the intercepts. In particular, the signs of $E[R|A=1]$ and $E[R|A=K]$ do not impact the monotonicity of the intercepts. For example, if $E[R|A=1]$ is the only non-positive one among all  $E[R|A=k], k=1,\cdots,K$, the first boundary $b_1$ would need to be at an extreme large value to prevent any subject from choosing $A=1$ as the optimal treatment. Such a scenario is not interesting in practice despite the fact that the monotonicity still holds with $b_{k}>b_{k+1}$ for $k=1,\cdots,K-2$.

\subsection{Excess 0-1 Risk and Excess $\phi-$Risk}

The following theorem shows that for any decision function $f$,
the excess risk of $f$ under the 0-1 loss, $\mathcal{R}(f)-\mathcal{R}(f^*)$,  can be bounded by the excess
risk of $f$ under the surrogate loss, $\mathcal{R}_{\phi}(f)-\mathcal{R}_{\phi}(f_{\phi}^*)$. 

\paragraph{Theorem 4.4}

For any measurable function $f:\mathcal{X}\rightarrow\mathbb{R}$
and any probability distribution of $(X,A,R)$, we have $\mathcal{R}_{\phi}(f)-\mathcal{R}_{\phi}(f_{\phi}^*) \geq \mathcal{R}(f)-\mathcal{R}(f^*) \geq 0$.

\paragraph{}

Because some of our theoretic discussions are based on the $\phi-$risk,
it is necessary to first show how the 0-1 loss risk $\mathcal{R}(f)$ could be controlled accordingly.
The proof of Theorem 4.4 uses the idea of partition and integration by dividing $\mathcal{R}_{\phi}(f)$ into $K-1$ parts with $\mathcal{R}_{\phi}(f)=\sum_{k=1}^{K-1}\mathcal{R}_{\phi}^{(k)}(f)$.  For each part $\mathcal{R}_{\phi}^{(k)}(f)$, we generalize the idea of \citet{zhao_estimating_2012} and make use of the risk bound theories in \citet{bartlett_convexity_2006} to derive the relationship between the two excess risks.

\subsection{Consistency and Convergence Rate}

Denote $\hat{f_n}$ as the sample solution for our proposed \GOWL\ as a minimizer of (\ref{eq:GOWL object}) with $f\in\mathcal{H}$. We next discuss the consistency of $\phi-$risk with $\hat{f_n}$ in the following Theorem 4.5.

\paragraph*{Theorem 4.5 (Consistency of $\mathcal{R}_{\phi}(\hat{f_n})$)}

Assume the tuning parameter $\lambda_{n}$ is selected such that $\lambda_{n}\rightarrow0$
and $n\lambda_{n}\rightarrow\infty$. Then for any distribution of $(X,A,R)$,
we have that ${ \mathcal{R}_{\phi}(\hat{f}_{n}) \rightarrow \underset{f\in\bar{\mathcal{H}}}{\inf}\mathcal{R}_{\phi}(f)}$ in probability as $n\rightarrow\infty$, where $\hat{f}_{n}$ is the empirical minimizer of (\ref{eq:GOWL object}) and $\bar{\mathcal{H}}$ denotes the closure of a selected space
$\mathcal{H}$.

\paragraph{}

By theorem 4.5, minimization of the $\phi-$risk depends on the selection
of $\mathcal{H}$. Additionally, if $f^{*}_{\phi}$, the global minimizer of (\ref{eq:surrogaterisk}), belongs to the
closure of $\underset{n\rightarrow\infty}{\lim\sup}\mathcal{H}$, where $\mathcal{H}$
could depend on $n$, then we have $\underset{f\in\bar{\mathcal{H}}}{\inf}\mathcal{R}_{\phi}(f)=\mathcal{R}_{\phi}(f^*_{\phi})$ and thus $\underset{n\rightarrow\infty}{\lim\inf}\mathcal{R}_{\phi}(\hat{f}_{n})=\mathcal{R}_{\phi}(f_{\phi}^*)$
in probability. This result will lead to $\underset{n\rightarrow\infty}{\lim\inf}\mathcal{R}(\hat{f}_{n})=\mathcal{R}(f^*)$
in probability by Theorem 4.4. In particular, the above conditions are met when $\mathcal{H}$
is an RKHS with the Gaussian kernel and the kernel bandwidth decreases
to zero as $n\rightarrow\infty$ (see \citet{zhao_estimating_2012} for a related discussion).

In the next theorem, we discuss the convergence rate of
the excess 0-1 risk $\mathcal{R}(\hat{f}_{n})-\mathcal{R}(f^*)$ based
on the geometric noise assumption for each measure $P^{(k)}$ introduced
in \citet{steinwart_fast_2007}. For our problem, we define the decision boundary for the optimal ITR as $\{2\eta(x^{(k)})-1=0\}$ in each classification subproblem between $\{1,\cdots,k\}$ and $\{k+1,\cdots,K\}$ for $k=1\cdots,K-1$, where{
\begin{eqnarray*}
\eta(x^{(k)})=\frac{E[R|X^{(k)}=x^{(k)}, A^{(k)}=1]-E[R|X^{(k)}=x^{(k)}, A^{(k)}=-1]}{E[R|X^{(k)}=x^{(k)}, A^{(k)}=1]+E[R|X^{(k)}=x^{(k)}, A^{(k)}=-1]}+\frac{1}{2}.
\end{eqnarray*}
}
Furthermore, we define the purity of the corresponding data set as $\Delta(x^{(k)})=|2\eta(x^{(k)})-1|$. Note that $\Delta(x^{(k)})$ can be viewed as a measure of closeness of $x^{(k)}$ to the corresponding $k$th decision boundary. Using these notations, we state the geometric noise assumption in our problem for each duplicate $k$ for $k=1,\cdots,K-1$ as follows: Let $X^{(k)}\in \mathbb{R}^{p}$ be compact, we define that the measure $\mathcal{P}^{k}$ has geometric noise exponent $q_k>0$ if there exists a constant $C_k>0$ such that $E[|2\eta(X^{(k)})-1|\exp(-\frac{\Delta(X^{(k)})^{2}}{t})]\leq C_k t^{q_k p/2}$, for $t>0$. According to \citet{steinwart_fast_2007}, the geometric noise exponent describes the concentration and the noise level of the data generating distribution near the decision boundary. As we will discuss further, the geometric noise exponent $q_k$ of the distribution of $(X^{(k)},A^{(k)},R^{(k)})$
depends on how the density of the data set decreases when the point gets close to the boundary. In the extreme case, $q_k$ can be arbitrarily large when $\eta(x^{(k)})$ is continuous and $\Delta(x^{(k)})>\delta>0$ for some constant $\delta>0$ (i.e., the distinctly separable case). In addition to the geometric noise condition, we also consider the RKHS associated with the Gaussian kernel as in \citet{steinwart_fast_2007} in Theorem 4.6. We use $\sigma_n$  to denote the kernel bandwidth for the Gaussian kernel.

\paragraph{Theorem 4.6 (Convergence Rate of the Excess Risk)}

Suppose that the distribution of $(X^{(k)},A^{(k)},R^{(k)})$
satisfies the geometric noise assumption with exponent $q_{k}\in(0,\infty)$
for $k=1,\cdots,K-1$.  Then for any $\delta>0$
and $\nu\in(0,2)$, there exists a $C$, which depends on $\nu,\delta$, the dimension $p$,
and the prior probability of the treatment $P(A|X)$, such that for $\forall\tau\geq1$ and $\sigma_{n}=\lambda_{n}^{-\frac{1}{(q+1)p}}$ for the Gaussian kernel,
we have $\text{Pr}^{*}(\mathcal{R}(\hat{f}_{n})\leq\mathcal{R}(f^*)+\epsilon)\geq1-e^{-\tau}$, where $q={\arg\max}_{q_k}{\lambda^{q_k/(q_{k}+1)}_n}$,  $\text{Pr}^{*}$ denotes outer probability and 
$\epsilon=C(\lambda_{n}^{-\frac{2}{2+\nu}+\frac{(2-\nu)(1+\delta)}{(2-\nu)(1+q)}}n^{-\frac{2}{2+\nu}}+\frac{\tau}{n\lambda_{n}}+\lambda_{n}^{\frac{q}{q+1}})$.


Taking a closer look at the $\epsilon$ expression in Theorem 4.6, we can find that the
first two terms can be treated as the bound for the stochastic error,
whereas the last term is an error bound for the noise associated with the corresponding RKHS. 
There is a trade off between the two components. For example, the
noise bound term will decrease and the stochastic error will inflate
if the RKHS is selected to be more complex. 
 Based on the $\epsilon$
expression, one can tell that an optimal choice of $\lambda_{n}$
is $n^{-\frac{2(1+q)}{(4+\nu)q+2+(2-\nu)(1+\delta)}}$ and the corresponding
rate of the excess risk can be expressed as $\mathcal{R}(\hat{f}_{n})-\mathcal{R}(f^*)\leq O_{p}(n^{-\frac{2q}{(4+\nu)q+2+(2-\nu)(1+\delta)}})$.
By the geometric noise exponent property, such $q$ can be sufficiently large
when different optimal treatment groups are separated well enough just as in the distinctly
separable case we discuss previously. Under this circumstance, the rate of convergence
can be almost $O_{p}\left(n^{-1/2}\right)$ when we let $\delta$
and $\nu$ be small. 


\section{Simulation Study}

In this section, we conduct simulation studies with both linear and non-linear ITR boundaries to assess the finite sample performance of the proposed \GOWL.
In both cases, we first generate a training set with the covariates $X_{1},\cdots,X_{p}$
from a uniform distribution $U\left(-1,1\right)$ and the treatment
$A$ from a discrete uniform distribution
ranging from $1$ to $K$, where $K=2,3,5$ and $7$ respectively. In each example, $X$ and $A$ are  independent. For each $K$,
we choose two training sample sizes to represent the small and
large sample scenarios. The reward $R$ follows $N(Q(X,A),1)$ with $Q(X,A)=\mu(X)+t(X,A)$, where $\mu(X)$ is the overall effect of $X$ and  $t(X,A)$ is the interaction that determines the
true optimal treatment. We maintain approximately 70\% of the generated rewards
as positive. For simplicity in simulation studies, except for the training set, we also generate an independent equal-size tuning set and a much larger testing set (10 times as large as the training set) with the same variables in each scenario. The tuning set is used to select the optimal tuning parameter $\lambda$ and the Gaussian kernel bandwidth $\sigma_n$. In particular, we choose $\lambda$ from $\{\frac{i}{n};i=0.1,1,10,100,500\}$ and $\sigma_{n}$ from $\{0.1,1,10\}$, where $n$ is the tuning size. The testing set is used to check the prediction performance of the models. For real data application, cross-validations are used for tuning parameter selection.

For comparisons, we manually modify some existing methods so that they can be used to detect the ITR for ordinal treatments. Specifically, we
pick OWL and $l_{1}$ penalized least squares including one way covariate-treatment interaction terms (PLS-$l_{1}$, \citet{qian_performance_2011})
to conduct a series of pairwise comparisons between $\{1,\cdots,k\}$ and $\{k+1,\cdots,K\}$ for $k=1,\cdots,K-1$. The final estimated
optimal treatment is obtained by summing through all pairwise prediction
results. For OWL, the original reward outcome is shifted to be all positive. For both
OWL and \GOWL, both the linear kernel (OWL-Linear
and \GOWL-Linear) and the Gaussian kernel (OWL-Gaussian and \GOWL-Gaussian)
are used for estimating the classifier. We select two criteria to evaluate the model performance: the misclassification rate (MISC),  and the MSE
of the value function (Value), i.e., the mean of squares of the difference between the Values under the estimated ITR versus under the optimal ITR for all replicates. Smaller values are preferred for both criteria
by definition. In particular, the first criterion measures the proportion
of correct treatment assignments. The second criterion is a more comprehensive
measure on how close the estimated ITR is to the true optimal ITR.
The value function estimate is defined as
$\mathbb{P}_{n}^{*}\left[\sum_{k=1}^{K-1}I\left(A^{(k)}=\mathcal{D}(X^{(k)})\right)R/P(A)\right]/\mathbb{P}_{n}^{*}\left[\sum_{k=1}^{K-1}I\left(A^{(k)}=\mathcal{D}(X^{(k)})\right)/P(A)\right]$,
where $\mathbb{P}_{n}^{*}$ denotes the empirical average of the
testing data set.

\subsection{Linear Boundary Examples}

We consider the following four scenarios with $\mu(X)$ and $t(X,A)$
defined as,
\begin{enumerate}
\item $K=2$: $\mu(X)=1+X_{1}+X_{2}+2X_{3}+0.5X_{4}$ and $t(X,A)=1.8\left(0.3-X_{1}-X_{2}\right)\left(2A-3\right);$
\item $K=3$: $\mu(X)=2+2X_{1}+X_{2}+0.5X_{3}$ and $t(X,A)=4\sum_{i=1}^{3}I\left(g(X)\in(b_{i-1},b_{i}]\right)(2-\left|A-i\right|)$,
where $g(X)=-X_{1}+2X_{2}+X_{3}+0.6X_{4}-1.5(X_{5}+X{}_{6})$, $b_{0}=-\infty$,
$b_{1}=-0.5$, $b_{2}=1$ and $b_{3}=\infty;$
\item $K=5$: $\mu(X)=2+2X_{1}+X_{2}+0.5X_{3}$ and $t(X,A)=4\sum_{i=1}^{5}I\left(g(X)\in(b_{i-1},b_{i}]\right)(2-\left|A-i\right|)$,
where $g(X)=-X_{1}+2X_{2}+X_{3}+0.6X_{4}-1.5(X_{5}+X{}_{6})$, $b_{0}=-\infty$,
$b_{1}=-1.9$, $b_{2}=-0.5$, $b_{3}=0.5$, $b_{4}=1.7$ and $b_{5}=\infty;$
\item $K=7$: $\mu(X)=2+2X_{1}+X_{2}+0.5X_{3}$ and $t(X,A)=4\sum_{i=1}^{7}I\left(g(X)\in(b_{i-1},b_{i}]\right)(2-\left|A-i\right|)$,
where $g(X)=-X_{1}+2X_{2}+X_{3}+0.6X_{4}-1.5(X_{5}+X{}_{6})$, $b_{0}=-\infty$,
$b_{1}=-2.1$, $b_{2}=-1.2$, $b_{3}=-0.4$, $b_{4}=0.4$, $b_{5}=1$,
$b_{6}=2.1$ and $b_{7}=\infty.$
\end{enumerate}
The simulated data set follows the assumption that the
true boundaries are parallel to each other. The 
cut-off values $b$ are set to encourage an evenly distributed true optimal
treatment from 1 to $K$ in samples.
Furthermore, the $t(X,A)$ functions are set to ensure that the reward outcome decreases symmetrically when the assigned treatment moves away from the optimal treatment towards high or low levels. The
training sample sizes are listed in Table \ref{tab:simu_linear}, which range from 30 to 500. We
repeat the simulation 50 times and present the prediction
results of the testing sets in Table \ref{tab:simu_linear}.

As shown in Table \ref{tab:simu_linear}, the proposed \GOWL\
reveals competitive accuracy rate in predicting ITR for testing data
sets in most of the cases. In general, when both the sample size $n$ and number
of treatment classes $K$ are small, the PLS-$l_{1}$ can be competitive because the true decision boundary is linear. However, when $K$ increases to 5 or 7, \GOWL\ outperforms all the other methods, especially in terms of the value function of the estimated ITR. Moreover, for the binary treatment with small $n$,
\GOWL\ performs comparable to PLS-$l_{1}$ whereas OWL shows relatively worse results with a larger MSE for the corresponding value function. When the number of treatment category $K$ increases, the
advantage of \GOWL\ becomes more significant in terms of both the misclassification
and value function comparisons. For example, \GOWL\ can maintain an average misclassification rate as 21\% even when $K$ increases to 7. One reason can be that the parallel decision boundary assumption of \GOWL\ matches the underlying truth and this can lead to robust estimate even when $K$ is large. Furthermore, under the true linear boundary cases, the performance of \GOWL\ with
the Gaussian kernel can be comparable to the case with the linear kernel when a proper tuning parameter is used. Thus a flexible nonparametric estimation procedure can be considered in practice when there is no prior knowledge about the shape of the underlying ITR boundaries. 

\subsection{Nonlinear Boundary Examples}

For the nonlinear boundary examples, we consider the following four scenarios with $\mu(X)$ and $t(X,A)$
defined as,
\begin{enumerate}
\item $K=2$: $\mu(X)=1+X_{1}^{2}+X_{2}^{2}-2X_{3}+0.5X_{4}$ and $t(X,A)=4(0.7-X_{1}^{2}-X_{2}^{2})(2A-3);$
\item $K=3$: $\mu(X)=2+2X_{1}+X_{2}+0.5X_{3}$ and $t(X,A)=4\sum_{i=1}^{3}I\left(g(X)\in(b_{i-1},b_{i}]\right)(2-\left|A-i\right|)$,
where $g(X)=-3-X_{1}^{2}+2\exp\{X_{2}\}+(X_{3}-0.6X_{4})^{2}+X_{5}^{3}+\exp\{X{}_{6}^{2}\}$,
$b_{0}=-\infty$, $b_{1}=0$, $b_{2}=1.3$ and $b_{3}=\infty;$
\item $K=5$: $\mu(X)=2+2X_{1}+X_{2}+0.5X_{3}$ and $t(X,A)=4\sum_{i=1}^{5}I\left(g(X)\in(b_{i-1},b_{i}]\right)(2-\left|A-i\right|)$,
where $g(X)=-3-X_{1}^{2}+2\exp\{X_{2}\}+(X_{3}-0.6X_{4})^{2}+X_{5}^{3}+\exp\{X{}_{6}^{2}\}$,
$b_{0}=-\infty$, $b_{1}=-0.4$, $b_{2}=0.3$, $b_{3}=1.1$, $b_{4}=2.1$
and $b_{5}=\infty;$
\item $K=7$: $\mu(X)=2+2X_{1}+X_{2}+0.5X_{3}$ and $t(X,A)=4\sum_{i=1}^{7}I\left(g(X)\in(b_{i-1},b_{i}]\right)(2-\left|A-i\right|)$,
where $g(X)=-3-X_{1}^{2}+2\exp\{X_{2}\}+(X_{3}-0.6X_{4})^{2}+X_{5}^{3}$,
$b_{0}=-\infty$, $b_{1}=-0.7$, $b_{2}=-0.2$, $b_{3}=0.4$, $b_{4}=1$,
$b_{5}=1.8$, $b_{6}=2.8$ and $b_{7}=\infty.$
\end{enumerate}
Similar to the linear boundary cases, we have a symmetric reward-treatment curve in each scenario. We repeat the simulation 50
times with the tuning parameters ranging in the same domain. The prediction results are displayed in Table \ref{tab:simu_nonlinear}.

From the results, none of the method performs well when the sample size is small because the true boundary
function has a complex structure. When $n$ becomes large, \GOWL\ with the Gaussian kernel outperforms PLS-$l_{1}$ in
all cases due to PLS-$l_{1}$'s wrong model specification. \GOWL\
with the Gaussian kernel shows better performance than OWL with the same kernel
in terms of both accuracy and value function error. For OWL, we find that the estimated optimal treatments are often the same as the actually assigned ones when $\sigma_{n}$ takes large values. 
This situation becomes more
severe when the treatment has seven categories. In addition, when $K=7$, we find that obtaining a low value function MSE becomes  challenging even for \GOWL\
with the Gaussian kernel. This may be due to the difficulty of the ITR detection for the ordinal treatments under nonlinear learning. Finally, we would like to note that the monotonic property of the intercept vectors $b$ holds in all  simulated cases above.

So far, our focus has been on examples with parallel boundaries.  We would like to point out that the proposed \GOWL\ could also work well when the parallel assumption of the true boundaries does not hold. Under these circumstances, one should consider using nonlinear learning techniques hence the estimated boundaries would be flexible enough to approach the underlying true boundaries. To illustrate the idea with a 2-dimensional graph, we use a  case with $n=300$, $p=2$ and $K=3$  and follow the previous settings to simulate $X$ and $A$. At this time, we have the Q-function generated by $Q(X,A,\mathcal{D}^*(X))=2+X_1+0.5X_2-2|A-\mathcal{D}^*(X)|$ where $\mathcal{D}^*(\cdot)$, the optimal treatment rule, is defined as, $\mathcal{D}^*(X)=1$ if $(X_1+1)^2+(X_2+1)^2<1$;  $\mathcal{D}^*(X)=2$ if $X_1+X_2>2/3$; $\mathcal{D}^*(X)=3$ otherwise. 

\begin{figure}[H]
\begin{centering}
\includegraphics[scale=0.35]{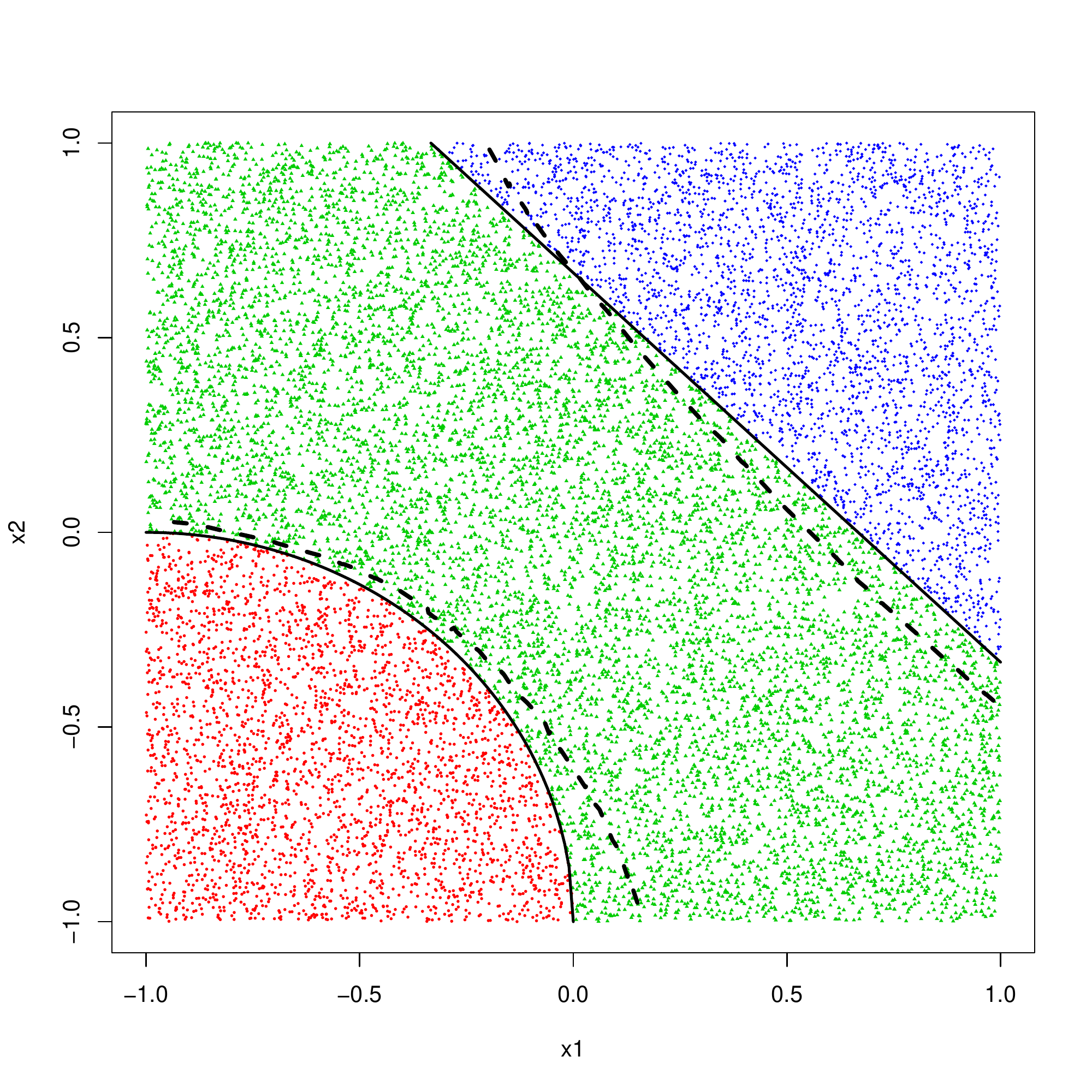}
\par\end{centering}

\protect\caption{Illustrating plot for the example with the true boundaries containing a linear line and a nonlinear curve. The solid curves indicate the true boundaries and the dashed curves represent the estimated boundaries by \GOWL-Gaussian in one simulation. The points correspond to the observations in the test set with the color representing the optimal treatment: red-1, green-2 and blue-3. \label{fig:Non-parallel}}
\end{figure}

Different from what were discussed in the previous examples, the current boundary set consists of a straight line and a one-fourth of a circle. Using \GOWL-Gaussian with the same tuning range as in Section 5.2, we plot the estimated boundaries (dashed curves) as well as the true boundaries (solid curves) in Figure \ref{fig:Non-parallel}. The results show that the estimated ITR could still capture the underlying pattern of the optimal ITR well since the RKHS with the Gaussian kernel is very flexible. We repeat the simulations for 50 times and the average testing misclassification rate is 5.05\%, which illustrates \GOWL's competitive prediction ability under the cases of complex boundaries.

\section{Dataset Applications}

We apply \GOWL\ to an irritable bowel syndrome clinical data set and a type 2 diabetes mellitus clinical observational study to assess its performance in real studies.

\subsection{Irritable Bowel Syndrome Dataset}

This dataset consists of a dose ranging
trial that aims to develop a treatment for irritable bowel syndrome
(IBS) (see \citet{biesheuvel_many--one_2002} for more details). The clinical
study enrolled four active treatment arms, corresponding to doses
1, 2, 3, 4 and placebo. The primary endpoint is a baseline adjusted
abdominal pain score with larger values corresponding to a better
treatment effect. There are 369 patients completing the study, with an
almost balanced allocation across the groups of different doses. The final data set only
contains three variables: patients' gender, treatment, and the adjusted
abdominal pain score. Approximately 72\% of the observed pain scores are greater than 0.

Given the small covariate dimension, we merge doses 1 and 2
together as the low dose group and merge doses 3 and 4 together as the high dose group. The average adjusted abdominal
pain scores of the total data set is 0.475, with standard deviation
equal to 0.769. To estimate the optimal ITR, we apply methods including PLS-$l_{1}$, OWL-Gaussian,
and GOWL-Gaussian, and modify the first two methods in the same way as 
in the simulation study. As to the evaluation criterion, we calculate the empirical
value function with the following cross-validation strategy. In particular, we randomly
partition the dataset into 5 equal-sized parts,  train the
model based on every 4 of them, and predict the value function using
the remaining part. We repeat the cross validation 50 times and summarize
the means and standard deviations of the predicted value function in Table \ref{tab:Real-Application:-Value}.

Table \ref{tab:Real-Application:-Value} shows that \GOWL\ returns the highest predicted value function with a moderately low standard deviation. By reassigning the treatment, GOWL could improve the predicted value function
by approximately 13\%. Furthermore, as to the estimated optimal treatment assignment, PLS-$l_{1}$
suggests the optimal treatment to be either placebo or low
dose. OWL assigns almost all the patients to the low dose group whereas
\GOWL\ suggests about 60\% patients in high dose and 40\% in low dose. In particular, around 70\% patients are female for those recommended to be in high dose group. This conclusion appears consistent to what \cite{biesheuvel_many--one_2002} reported. 

\subsection{Type 2 Diabetes Mellitus Clinical Observational Study}

In this section, we apply the proposed method to a type 2 diabetes mellitus (T2DM) observational study to assess its performance in real life data application. This study includes people with T2DM during 2012-2013, from clinical practice research datalink (CPRD)1 (Herrett 2015). Three anti-diabetic therapies have been considered in this study: glucagon-like peptide-1 (GLP-1) receptor agonist, long-acting insulin only, and a regime including short-acting insulin. The primary target variable is the change of HbA1c before and after the treatment, and seven clinical factors are used including age, gender, ethnicity, body mass index, high-density lipoprotein cholesterol (HDL), low-density lipoprotein cholesterol (LDL) and smoking status. In total, 634 patients satisfying aforementioned requirements are while around 5\% have complete observations.

To handle the missing data issue before analysis, we first remove all
the covariates with missing proportions greater than 70\%. Then, we conduct
a $t$ test for each covariate to detect whether its missing pattern
impacts the mean of outcome significantly. According to the Bonferroni
multiple-testing adjusted $p$ value, we remove all of the covariates with
insignificant test results. For the continuous variables with significant test $p$ values, we remove all of their incomplete observations. For categorical covariates having significant test results, we relabel the missing value as a new class when encoding the covariate. After the data preprocessing, there are 10 covariates with 142 observations in total.

Similar to the previous analysis, we apply PLS-$l_{1}$, OWL-Linear,
OWL-Gaussian, \GOWL-Linear, and \GOWL-Gaussian to estimate the ITR with
the first three methods modified in the same way. We use the
inverse value of the HbA1c change as the reward in estimating the ITR
since a smaller HbA1c is desired. In order to obtain the propensity score $P(A|X)$ before using OWL and \GOWL, we fit an ordinal logistic model with the cleaned data set using the treatment as the response and all 10 covariates as predictors. As to the criterion, similar to the irritable bowel syndrome example, we calculate the predicted value function using the same formula as in the simulation study over 50 replications of 5-fold cross-validation. Table \ref{tab:Real-Application_T2D} summarizes the
means and standard deviations of the empirical value function from the
training and validation sets.

To further demonstrate how much improvement the proposed method obtain, we also calculate
the value function with the original treatments and
the average value function with treatment being randomly assigned 50 times.
The empirical means of the value functions are 2.205 and 2.104 with the standard deviation for the random assignment to be 0.131.

According to Table \ref{tab:Real-Application_T2D}, \GOWL\
achieves both the highest mean and the lowest
standard deviation of the empirical value function in the prediction results. In addition, the
three linear models are outperformed by the nonlinear models possibly due
to their suboptimal model specification for this application. As to the distribution
of estimated optimal treatment assignments, the PLS-$l_{1}$ only includes 
long-acting insulin as the optimal treatment. OWL-Gaussian chooses approximately 83\% of the
patients to be in either the GLP-1 group or short-acting insulin group.
\GOWL-Gaussian assigns approximately 50\% patients into the short-acting
insulin group while assigning the rest into one of the other two groups in
a more even way.

\section{Conclusion}

In this paper, we use a modified loss function to improve the performance
of OWL and then generalize OWL to solve the ordinal
treatment problems. In particular, the proposed \GOWL\ converts the optimal
ordinal treatment finding problem into multiple optimal binary treatment
finding subproblems under certain restrictions. The estimating process
produces a group of estimated optimal treatment
boundaries which would never cross and have monotonic intercepts. Such boundaries can make the ITR estimates more stable and interpretable
in practice. 


There are various possible extensions for \GOWL\ that could be considered.
For example, one can incorporate a variable selection component into the objective function. In the literature, \citet{xu_regularized_2015} proposed variable selection in the linear case and \citet{zhou_residual_2015} extended the idea for kernel learning. According to their ideas, one nature extension for \GOWL\ is to include an $l_{1}$ penalty of the parameters into its optimization problem. In this way, variable sparsity could be achieved simultaneously when detecting the optimal ITR. The second possible extension that might improve the performance of \GOWL\ is to modify the outcome in its optimization problem which is originally the reward $R$. Specifically, according to \citet{fu_estimating_2016} and \citet{zhao_new_2015}, one can consider fitting a model with $R$ versus $X$ and then put the residuals as the outcome in the optimization problem of \GOWL\ instead. Such an adjustment is likely to further improve the ITR estimation results for some finite sample scenarios. Another potential extension is to apply \GOWL\ to solve the dynamic treatment regime problem, i.e. how to maximize the clinical rewards when there are multiple stages of treatments. The idea of \citet{zhao_new_2015} could possibly be adapted to such developments.

\bibliographystyle{apa-good}

\newpage
\pagenumbering{gobble}

\begin{table}[H]
\begin{centering}
\begin{tabular}{cccccccccccc}
\hline 
\multicolumn{2}{c}{{\footnotesize{}Methods}} & \multicolumn{2}{c}{{\footnotesize{}PLS-$l_{1}$}} & \multicolumn{2}{c}{{\footnotesize{}OWL-Linear}} & \multicolumn{2}{c}{{\footnotesize{}OWL-Gaussian}} & \multicolumn{2}{c}{{\footnotesize{}GOWL-Linear}} & \multicolumn{2}{c}{{\footnotesize{}GOWL-Gaussian}}\tabularnewline
\hline 
{\footnotesize{}$K$} & {\footnotesize{}$n$} & {\footnotesize{}MISC} & {\footnotesize{}Value} & {\footnotesize{}MISC} & {\footnotesize{}Value} & {\footnotesize{}MISC} & {\footnotesize{}Value} & {\footnotesize{}MISC} & {\footnotesize{}Value} & {\footnotesize{}MISC} & {\footnotesize{}Value}\tabularnewline
\hline 
\multirow{4}{*}{{\footnotesize{}2}} & \multirow{2}{*}{{\footnotesize{}30}} & \textbf{\footnotesize{}0.117} & \textbf{\footnotesize{}0.128} & {\footnotesize{}0.198} & {\footnotesize{}0.464} & {\footnotesize{}0.196} & {\footnotesize{}0.454} & {\footnotesize{}0.155} & {\footnotesize{}0.166} & {\footnotesize{}0.122} & {\footnotesize{}0.138}\tabularnewline
 &  & {\footnotesize{}(0.107)} & {\footnotesize{}(0.111)} & {\footnotesize{}(0.168)} & {\footnotesize{}(0.327)} & {\footnotesize{}(0.148)} & {\footnotesize{}(0.290)} & {\footnotesize{}(0.121)} & {\footnotesize{}(0.133)} & {\footnotesize{}(0.087)} & {\footnotesize{}(0.145)}\tabularnewline
 & \multirow{2}{*}{{\footnotesize{}300}} & {\footnotesize{}0.130} & {\footnotesize{}0.018} & {\footnotesize{}0.055} & {\footnotesize{}0.081} & {\footnotesize{}0.105} & {\footnotesize{}0.084} & {\footnotesize{}0.077} & {\footnotesize{}0.014} & \textbf{\footnotesize{}0.032} & \textbf{\footnotesize{}0.012}\tabularnewline
 &  & {\footnotesize{}(0.045)} & {\footnotesize{}(0.005)} & {\footnotesize{}(0.024)} & {\footnotesize{}(0.054)} & {\footnotesize{}(0.073)} & {\footnotesize{}(0.036)} & {\footnotesize{}(0.034)} & {\footnotesize{}(0.009)} & {\footnotesize{}(0.011)} & {\footnotesize{}(0.006)}\tabularnewline
\hline 
\multirow{4}{*}{{\footnotesize{}3}} & \multirow{2}{*}{{\footnotesize{}30}} & {\footnotesize{}0.269} & {\footnotesize{}0.450} & {\footnotesize{}0.425} & {\footnotesize{}0.620} & {\footnotesize{}0.422} & {\footnotesize{}0.633} & \textbf{\footnotesize{}0.220} & \textbf{\footnotesize{}0.270} & {\footnotesize{}0.235} & {\footnotesize{}0.273}\tabularnewline
 &  & {\footnotesize{}(0.152)} & {\footnotesize{}(0.288)} & {\footnotesize{}(0.349)} & {\footnotesize{}(0.413)} & {\footnotesize{}(0.350)} & {\footnotesize{}(0.315)} & {\footnotesize{}(0.150)} & {\footnotesize{}(0.198)} & {\footnotesize{}(0.157)} & {\footnotesize{}(0.118)}\tabularnewline
 & \multirow{2}{*}{{\footnotesize{}300}} & {\footnotesize{}0.285} & {\footnotesize{}0.044} & {\footnotesize{}0.261} & {\footnotesize{}0.398} & {\footnotesize{}0.243} & {\footnotesize{}0.468} & \textbf{\footnotesize{}0.032} & \textbf{\footnotesize{}0.028} & {\footnotesize{}0.055} & {\footnotesize{}0.029}\tabularnewline
 &  & {\footnotesize{}(0.071)} & {\footnotesize{}(0.019)} & {\footnotesize{}(0.165)} & {\footnotesize{}(0.271)} & {\footnotesize{}(0.176)} & {\footnotesize{}(0.364)} & {\footnotesize{}(0.021)} & {\footnotesize{}(0.012)} & {\footnotesize{}(0.043)} & {\footnotesize{}(0.013)}\tabularnewline
\hline 
\multirow{4}{*}{{\footnotesize{}5}} & \multirow{2}{*}{{\footnotesize{}50}} & {\footnotesize{}0.608} & {\footnotesize{}0.616} & {\footnotesize{}0.589} & {\footnotesize{}0.878} & {\footnotesize{}0.355} & {\footnotesize{}0.758} & {\footnotesize{}0.351} & {\footnotesize{}0.290} & \textbf{\footnotesize{}0.337} & \textbf{\footnotesize{}0.267}\tabularnewline
 &  & {\footnotesize{}(0.241)} & {\footnotesize{}(0.432)} & {\footnotesize{}(0.330)} & {\footnotesize{}(0.320)} & {\footnotesize{}(0.329)} & {\footnotesize{}(0.345)} & {\footnotesize{}(0.256)} & {\footnotesize{}(0.175)} & {\footnotesize{}(0.229)} & {\footnotesize{}(0.145)}\tabularnewline
 & \multirow{2}{*}{{\footnotesize{}500}} & {\footnotesize{}0.436} & {\footnotesize{}0.272} & {\footnotesize{}0.303} & {\footnotesize{}0.305} & {\footnotesize{}0.344} & {\footnotesize{}0.295} & {\footnotesize{}0.163} & {\footnotesize{}0.042} & \textbf{\footnotesize{}0.118} & \textbf{\footnotesize{}0.030}\tabularnewline
 &  & {\footnotesize{}(0.122)} & {\footnotesize{}(0.129)} & {\footnotesize{}(0.263)} & {\footnotesize{}(0.319)} & {\footnotesize{}(0.184)} & {\footnotesize{}(0.283)} & {\footnotesize{}(0.095)} & {\footnotesize{}(0.033)} & {\footnotesize{}(0.095)} & {\footnotesize{}(0.018)}\tabularnewline
\hline 
\multirow{4}{*}{{\footnotesize{}7}} & \multirow{2}{*}{{\footnotesize{}50}} & {\footnotesize{}0.672} & {\footnotesize{}1.609} & {\footnotesize{}0.707} & {\footnotesize{}0.910} & {\footnotesize{}0.721} & {\footnotesize{}1.625} & \textbf{\footnotesize{}0.414} & {\footnotesize{}0.404} & {\footnotesize{}0.420} & \textbf{\footnotesize{}0.375}\tabularnewline
 &  & {\footnotesize{}(0.327)} & {\footnotesize{}(0.855)} & {\footnotesize{}(0.317)} & {\footnotesize{}(0.480)} & {\footnotesize{}(0.303)} & {\footnotesize{}(0.575)} & {\footnotesize{}(0.282)} & {\footnotesize{}(0.244)} & {\footnotesize{}(0.290)} & {\footnotesize{}(0.308)}\tabularnewline
 & \multirow{2}{*}{{\footnotesize{}500}} & {\footnotesize{}0.587} & {\footnotesize{}0.371} & {\footnotesize{}0.491} & {\footnotesize{}0.364} & {\footnotesize{}0.522} & {\footnotesize{}0.365} & \textbf{\footnotesize{}0.210} & \textbf{\footnotesize{}0.098} & {\footnotesize{}0.227} & {\footnotesize{}0.103}\tabularnewline
 &  & {\footnotesize{}(0.247)} & {\footnotesize{}(0.280)} & {\footnotesize{}(0.247)} & {\footnotesize{}(0.282)} & {\footnotesize{}(0.179)} & {\footnotesize{}(0.219)} & {\footnotesize{}(0.161)} & {\footnotesize{}(0.072)} & {\footnotesize{}(0.145)} & {\footnotesize{}(0.040)}\tabularnewline
\hline
\end{tabular}
\par\end{centering}

\protect\caption{Results of linear boundary examples: $K$ represents the number of
treatment levels; $n$ represents the training set size; the MISC column
gives the mean and standard deviation of the misclassification rate; and the Value column gives
the mean and standard deviation of the value function MSE. PLS$-l_1$ represents penalized least squares including covariate-treatment interactions with $l_1$ penalty \citep{qian_performance_2011}; OWL represents the outcome weighted learning and GOWL represents the proposed generalized outcome weighted learning. In each scenario, the model producing the best criterion is in bold. \label{tab:simu_linear}}
\end{table}

\begin{table}[H]
\begin{centering}
\begin{tabular}{cccccccccccc}
\hline 
\multicolumn{2}{c}{{\footnotesize{}Methods}} & \multicolumn{2}{c}{{\footnotesize{}PLS-$l_{1}$}} & \multicolumn{2}{c}{{\footnotesize{}OWL-Linear}} & \multicolumn{2}{c}{{\footnotesize{}OWL-Gaussian}} & \multicolumn{2}{c}{{\footnotesize{}GOWL-Linear}} & \multicolumn{2}{c}{{\footnotesize{}GOWL-Gaussian}}\tabularnewline
\hline 
{\footnotesize{}$K$} & {\footnotesize{}$n$} & {\footnotesize{}MISC} & {\footnotesize{}Value} & {\footnotesize{}MISC} & {\footnotesize{}Value} & {\footnotesize{}MISC} & {\footnotesize{}Value} & {\footnotesize{}MISC} & {\footnotesize{}Value} & {\footnotesize{}MISC} & {\footnotesize{}Value}\tabularnewline
\hline 
\multirow{4}{*}{{\footnotesize{}2}} & \multirow{2}{*}{{\footnotesize{}30}} & {\footnotesize{}0.496} & {\footnotesize{}2.107} & {\footnotesize{}0.412} & {\footnotesize{}1.791} & \textbf{\footnotesize{}0.353} & \textbf{\footnotesize{}1.301} & {\footnotesize{}0.438} & {\footnotesize{}1.846} & {\footnotesize{}0.423} & {\footnotesize{}1.580}\tabularnewline
 &  & {\footnotesize{}(0.130)} & {\footnotesize{}(0.366)} & {\footnotesize{}(0.086)} & {\footnotesize{}(0.574)} & {\footnotesize{}(0.091)} & {\footnotesize{}(0.580)} & {\footnotesize{}(0.074)} & {\footnotesize{}(0.300)} & {\footnotesize{}(0.069)} & {\footnotesize{}(0.548)}\tabularnewline
 & \multirow{2}{*}{{\footnotesize{}300}} & {\footnotesize{}0.396} & {\footnotesize{}1.983} & {\footnotesize{}0.374} & {\footnotesize{}1.815} & {\footnotesize{}0.184} & {\footnotesize{}0.110} & {\footnotesize{}0.339} & {\footnotesize{}1.510} & \textbf{\footnotesize{}0.089} & \textbf{\footnotesize{}0.015}\tabularnewline
 &  & {\footnotesize{}(0.08)} & {\footnotesize{}(0.134)} & {\footnotesize{}(0.076)} & {\footnotesize{}(0.357)} & {\footnotesize{}(0.06)} & {\footnotesize{}(0.096)} & {\footnotesize{}(0.045)} & {\footnotesize{}(0.438)} & {\footnotesize{}(0.024)} & {\footnotesize{}(0.005)}\tabularnewline
\hline 
\multirow{4}{*}{{\footnotesize{}3}} & \multirow{2}{*}{{\footnotesize{}30}} & {\footnotesize{}0.461} & {\footnotesize{}1.191} & {\footnotesize{}0.470} & {\footnotesize{}2.640} & {\footnotesize{}0.468} & {\footnotesize{}1.574} & {\footnotesize{}0.403} & {\footnotesize{}1.214} & \textbf{\footnotesize{}0.370} & \textbf{\footnotesize{}0.909}\tabularnewline
 &  & {\footnotesize{}(0.225)} & {\footnotesize{}(0.347)} & {\footnotesize{}(0.107)} & {\footnotesize{}(0.538)} & {\footnotesize{}(0.106)} & {\footnotesize{}(0.608)} & {\footnotesize{}(0.094)} & {\footnotesize{}(0.445)} & {\footnotesize{}(0.066)} & {\footnotesize{}(0.218)}\tabularnewline
 & \multirow{2}{*}{{\footnotesize{}300}} & {\footnotesize{}0.345} & {\footnotesize{}0.645} & {\footnotesize{}0.361} & {\footnotesize{}1.495} & {\footnotesize{}0.362} & {\footnotesize{}0.861} & {\footnotesize{}0.224} & {\footnotesize{}0.403} & \textbf{\footnotesize{}0.146} & \textbf{\footnotesize{}0.048}\tabularnewline
 &  & {\footnotesize{}(0.18)} & {\footnotesize{}(0.239)} & {\footnotesize{}(0.092)} & {\footnotesize{}(0.527)} & {\footnotesize{}(0.089)} & {\footnotesize{}(0.448)} & {\footnotesize{}(0.08)} & {\footnotesize{}(0.136)} & {\footnotesize{}(0.04)} & {\footnotesize{}(0.018)}\tabularnewline
\hline 
\multirow{4}{*}{{\footnotesize{}5}} & \multirow{2}{*}{{\footnotesize{}50}} & {\footnotesize{}0.578} & {\footnotesize{}0.690} & {\footnotesize{}0.642} & {\footnotesize{}1.586} & {\footnotesize{}0.624} & {\footnotesize{}2.020} & \textbf{\footnotesize{}0.521} & {\footnotesize{}1.059} & {\footnotesize{}0.525} & \textbf{\footnotesize{}0.950}\tabularnewline
 &  & {\footnotesize{}(0.226)} & {\footnotesize{}(0.519)} & {\footnotesize{}(0.483)} & {\footnotesize{}(0.713)} & {\footnotesize{}(0.179)} & {\footnotesize{}(1.073)} & {\footnotesize{}(0.124)} & {\footnotesize{}(0.588)} & {\footnotesize{}(0.109)} & {\footnotesize{}(0.326)}\tabularnewline
 & \multirow{2}{*}{{\footnotesize{}500}} & {\footnotesize{}0.548} & {\footnotesize{}0.316} & {\footnotesize{}0.468} & {\footnotesize{}1.812} & {\footnotesize{}0.396} & {\footnotesize{}1.348} & {\footnotesize{}0.412} & {\footnotesize{}0.358} & \textbf{\footnotesize{}0.246} & \textbf{\footnotesize{}0.185}\tabularnewline
 &  & {\footnotesize{}(0.28)} & {\footnotesize{}(0.028)} & {\footnotesize{}(0.149)} & {\footnotesize{}(1.035)} & {\footnotesize{}(0.133)} & {\footnotesize{}(0.384)} & {\footnotesize{}(0.193)} & {\footnotesize{}(0.078)} & {\footnotesize{}(0.119)} & {\footnotesize{}(0.136)}\tabularnewline
\hline 
\multirow{4}{*}{{\footnotesize{}7}} & \multirow{2}{*}{{\footnotesize{}50}} & {\footnotesize{}0.727} & {\footnotesize{}3.489} & {\footnotesize{}0.707} & {\footnotesize{}4.172} & {\footnotesize{}0.716} & {\footnotesize{}2.412} & {\footnotesize{}0.590} & {\footnotesize{}0.695} & \textbf{\footnotesize{}0.563} & \textbf{\footnotesize{}0.503}\tabularnewline
 &  & {\footnotesize{}(0.319)} & {\footnotesize{}(0.989)} & {\footnotesize{}(0.578)} & {\footnotesize{}(0.923)} & {\footnotesize{}(0.266)} & {\footnotesize{}(0.685)} & {\footnotesize{}(0.178)} & {\footnotesize{}(0.561)} & {\footnotesize{}(0.163)} & {\footnotesize{}(0.388)}\tabularnewline
 & \multirow{2}{*}{{\footnotesize{}500}} & {\footnotesize{}0.665} & {\footnotesize{}2.754} & {\footnotesize{}0.722} & {\footnotesize{}1.757} & {\footnotesize{}0.541} & {\footnotesize{}1.414} & {\footnotesize{}0.610} & {\footnotesize{}1.378} & \textbf{\footnotesize{}0.445} & \textbf{\footnotesize{}0.795}\tabularnewline
 &  & {\footnotesize{}(0.287)} & {\footnotesize{}(0.798)} & {\footnotesize{}(0.238)} & {\footnotesize{}(0.424)} & {\footnotesize{}(0.21)} & {\footnotesize{}(0.253)} & {\footnotesize{}(0.244)} & {\footnotesize{}(0.146)} & {\footnotesize{}(0.168)} & {\footnotesize{}(0.17)}\tabularnewline
\hline 
\end{tabular}
\par\end{centering}

\protect\caption{Results of nonlinear boundary examples: $K$ represents the number
of treatment levels, $n$ represents the training set size, MISC column
gives the mean and standard deviation of the misclassification rate and Value column gives
the mean and standard deviation of the value function MSE\label{tab:simu_nonlinear}}
\end{table}

\begin{table}[H]
\begin{centering}
\begin{tabular}{ccc}
\hline 
Model & Training & Testing\tabularnewline
\hline 
PLS-$l_{1}$ & 2.257 (0.001) & 2.206 (0.059)\tabularnewline 
OWL-Linear & 2.335 (0.017) & 2.305 (0.072)\tabularnewline 
OWL-Gaussian & 2.456 (0.011) & 2.285 (0.049)\tabularnewline 
GOWL-Linear & 2.378 (0.047) & 2.332 (0.095)\tabularnewline 
GOWL-Gaussian & \textbf{2.486 (0.025)} & \textbf{2.383 (0.060)}\tabularnewline
\hline 
\end{tabular}
\par\end{centering}

\protect\caption{Analysis Results for the T2DM Dataset. Empirical Value Function Results using 5-fold Cross-Validation with 50 Replications are reported\label{tab:Real-Application_T2D}. For
comparison, the original assigned treatment strategy has the value
function 2.205 and the randomly assigned treatment method has average value
function 2.104 in testing sets with standard deviation 0.131.}
\end{table}

\begin{table}[H]
\begin{centering}
\begin{tabular}{cccc}
\hline 
Methods & PLS-$l_{1}$ & OWL-Gaussian & GOWL-Gaussian\tabularnewline
\hline 
Mean (Std) & 0.491 (0.029) & 0.503 (0.004) & 0.537 (0.011)\tabularnewline
\hline 
\end{tabular}
\par\end{centering}

\protect\caption{Empirical Value Function Results using 5-fold Cross-Validation for the IBS dataset \label{tab:Real-Application:-Value}}
\end{table}

\newpage

\end{document}